\documentclass[useAMS,usenatbib]{mn2e}
\usepackage{graphicx}

\def \be {\begin{equation}}
\def \ee {\end{equation}} 
\def \bdm {\begin{displaymath}}
\def \edm {\end{displaymath}}
\def \bea {\begin{eqnarray}}
\def \eea {\end{eqnarray}}

\title[Early evolution of newly born magnetars with a strong toroidal field]
{Early evolution of newly born magnetars with a strong toroidal field} 
\author[S. Dall'Osso, S.N. Shore, L. Stella]
{S. Dall'Osso$^{1,2}$, S. N. Shore$^{1}$ and L. Stella$^{2}$\\
$^{1}$ Universit\`{a} degli Studi di Pisa, Dipartimento di Fisica ``E. Fermi'' 
\& INFN, Largo B. Pontecorvo 3, Pisa, Italy \\
$^{2}$ INAF - Osservatorio Astronomico di Roma, via di 
Frascati 33, 00040 Monteporzio Catone (Roma)}
\begin{document}

\date{November 20 2008}

\maketitle

\label{firstpage}

\begin{abstract}
We present a state-of-the-art scenario for newly born magnetars as 
strong sources of Gravitational Waves (GWs)in the early days after formation.
We address several aspects of the astrophysics of rapidly rotating,
ultramagnetized neutron stars (NSs), including early cooling before transition 
to superfluidity, the effects of the magnetic field on the equilibrium shape of 
NSs, the internal dynamical state of a fully degenerate, oblique rotator 
and the strength of the electromagnetic torque on the newly born NS. We show 
that our scenario is consistent with recent 
studies of SNRs surrounding AXPs and SGRs in the Galaxy that constrain the
electromagnetic energy input from the central NS to be $\leq 10^{51}$ erg 
\citep{ViKu}. We further show that if this condition is met, then the GW 
signal from such sources is potentially detectable with the forthcoming 
generation of GW detectors up to Virgo cluster distances 
where an event rate $\sim$ 1/yr can be estimated (Stella et al. 2005). 
Finally, we point out that the decay of an internal magnetic field in the 
$10^{16}$ G range couples strongly to the NS cooling at very early stages, 
thus significantly slowing down both processes: the field can remain 
this strong for at least $10^3$ yrs, during which the core temperature stays
higher than several $\times 10^8$ K.

\end{abstract} 

\section{Introduction}
\label{intro}
Gravitational wave emission from spheroidal, rapidly rotating, isolated 
NSs has long been considered in the astrophysical literature 
(\citealt{OstrGun}). A natural origin of this distortion can 
be the anisotropic pressure from the internal magnetic field. However, before 
the early nineties, the inferred magnetic fields of NSs were $\sim 10^{13}$ 
G, which implied tiny deviations from spherical symmetry and, thus, 
weak GW emission that would be detectable, in principle, only from the nearest 
sources and with years-long observations \citep{BoGou}.\\ 
The discovery of the Soft Gamma Ray Repeaters (SGRs) and Anomalous X-ray 
Pulsars (AXPs) (cfr. \citealt{Maz}, \citealt{MeSte}) led to the idea 
that these peculiar high-energy sources could be ultramagnetized NSs, 
magnetars, with external (dipole) fields in the $10^{14} \div 10^{15}$ G range 
and with internal fields at least one order of magnitude stronger 
(\citealt{Pac}, Duncan \& Thompson 1992, DT92 hereafter, Thompson \& Duncan 
1993, 1995, 1996, hereafter \citealt{TD93, TD95, TD96}, respectively). 
The core of the proto-neutron star (PNS) experiences a phase of 
neutrino-driven turbulent convection (during a few tens of seconds). Strong 
differential rotation is also present if the nascent NS is spinning at 
millisecond period (we indicate the initial spin, in milliseconds, as 
$\mbox{\small{P}}_{\mbox{\tiny{i}}}$/ms), providing a total free energy reservoir 
of up to $\sim 10^{52} (\mbox{\small{P}}_{\mbox{\tiny{i}}}/ \mbox{\small{ms}})^{-2}$ 
erg. The combination of these two factors can power an $\alpha-\Omega$ dynamo 
that acts coherently over the whole NS core \citep{DT92}. 
A seed magnetic field is twisted into a mainly toroidal configuration and its 
intensity amplified to values $\sim 10^{16}$ G. The total energy available  
from the core differential rotation corresponds to a maximum field of 
$\sim 10^{17}$ G \citep{TD93, Du98}. Millisecond spin periods are required for 
differential rotation to play an important role and to achieve magnetic field
coherence lengthscales comparable to the stellar radius. In more slowly 
rotating NSs differential rotation is almost negligible and a 
convection-driven, $\alpha$-type dynamo results that acts stochastically and 
leads to a much weaker large scale field \citep{DT92, TD93}. 
In this formation scenario, magnetars represent those NSs that are born very 
rapidly rotating, at $\leq$ 3 ms spin period. \\
As the magnetar model received increasing support from observations during the 
last decade it was shortly after realized that such objects at birth represent 
promising sources of GWs, because of their large magnetic deformation and 
rapid rotation \citep{Io01, Pal}. These early suggestions assumed that a simple 
dipole field extended throughout the NS and its magnetosphere thus causing NS 
deformation and magnetodipole spindown. 
\Citet{Cut02} first highlighted the crucial role of the internal field 
structure in the GW emission efficiency of a magnetically distorted NS. He 
pointed out that a NS with a strong internal toroidal field has a prolate 
shape and would thus provide the best chance of being a strong GW emitter. 
Conversely, GW emission from a NS with an oblate distortion (such as that 
caused by a dipole field) would be rapidly quenched, since such an object 
would tend to spin around its symmetry axis (see $\S$ \ref{earlyspindown}).
%
Based on the energetics and likely recurrence time of Giant Flares such as the 
Dec 27 event from SGR 1806-20, \citet{Ste05} derived a lower limit of $\sim 8 
\times 10^{15}$ G for the strength of the core magnetic field in this object. 
This suggests that internal fields of magnetars may be even stronger than 
previously thought, reaching the $10^{16}$ G range. \citet{Laz}, \citet{Nak} 
and \citet{PoSt}, based on the apparent lack of  Giant Flare-like events in 
the BATSE archive, suggest longer recurrence times for such events, thus 
reducing the minimum energy requirement proposed by \citet{Ste05}.
However, the correlation between short GRBs and local galaxies found by 
\citet{Tanv} leaves room for a fraction of such events to be associated with
a population of young NSs. Furthermore, \citet{Kam07} showed 
that, if the thermal emission observed from AXPs is indeed powered by the 
decay of a magnetar's B-field, then the field strength in the NS crust must 
be 
$\sim 10^{16}$ G.\\
Fields $\sim 10^{16}$ G would imply remarkably large (prolate) magnetic 
deformations, $\epsilon_B \sim 10^{-3}$
(see $\S$ \ref{ellipticity}) which, 
in turn, would make GW emission at birth from such objects strong enough to be 
detectable up to the Virgo cluster distance, with Advanced LIGO/Virgo-class 
detectors. The integrated magnetar formation rate in Virgo is estimated to 
be $\sim$ 1 per year (\citealt{Ste05}), as opposed to $\sim 10^{-3}$ per 
year in the Galaxy alone (\citealt{Gae99}), thus making newly formed 
magnetars potentially interesting sources for next generation GW detectors, 
with a fairly high rate of occurence.\\
As an alternative scenario, it was proposed that strongly magnetic NSs 
can form as a direct result of magnetic flux conservation during 
core-collapse of a massive star with unusually strong magnetic field 
(\citealt{FerWick}, cf. \citealt{Uso}). In this picture, magnetars need not
 be born rapidly rotating: their magnetic field is stronger than in ordinary 
NSs because the magnetic dipole moment of their progenitor 
stars was particularly large. \citet{FerWick} have shown that a population 
of NSs with dipole fields up to $\sim 10^{15}$ G can be obtained in this 
framework starting from a realistic Galactic population of 
high-mass, high-field stars. We note, however, that the upper end of their 
field distribution seems to fall short of the minimal requirement within the 
magnetar model. Internal fields of \textit{at least} several times $10^{15}$ G 
are needed both to explain the overall energy output of SGRs and AXPs 
(\citealt{TD95}, Thompson \& Duncan 2001, hereafter \citealt{TD01}), 
and to make the magnetic field decay timescale comparable to, or shorter than, 
the estimated ages ($\sim 10^4$ yrs) of these sources \citep{TD96}.\\ 
\citet{GepRe} have shown that a field amplified to 
magnetar-strength in a newly formed NS will survive an early unstable phase 
(and subsequent dissipation) only if the NS spin period is shorter than $\sim$ 
5 ms, thus reinforcing the case for fast spins at birth.\\
If magnetars are born with millisecond spin periods (hereafter millisecond 
magnetar), their initial spin energy will be $E_{\mbox{\tiny{spin}}} \sim 3 \times 
10^{52} (\mbox{\small{P}}_{\mbox{\tiny{i}}} / \mbox{\small{ms}})^{-2}$ erg. Here and
throughout we assume a typical radius of 12 km and a mass of 1.4 M$_{\odot}$. 
The spindown timescale through magnetic dipole radiation will be very short, 
$\sim$  one day for external dipole fields in the $10^{14}$ G range. Most of 
the initial spin energy would thus be rapidly transferred to the surrounding 
supernova ejecta \citep{AlHo}. Recent studies have also shown that, 
for newly formed magnetars with dipole fields in excess of $\sim (6\div 7) 
\times 10^{14}$ G and spin period of $\approx$ 1 ms, strongly magnetized, 
relativistic winds are produced which can carry away most of the 
initial spin energy in a matter of minutes, thus opening the possibility of an 
even faster transfer of energy to the ejecta \citep{Tho04, Buc06, Metz07}.
In both cases, present-day SNRs around known magnetars should bear the 
signature of this larger-than-usual energy injection. For initial spin periods 
less than 3 ms, the injected energy would be $> 3 \times 10^{51}$ erg, making 
these remnants significantly more energetic than those surrounding ordinary 
NSs.\\
The X-ray spectra of the SNRs surrounding known magnetar candidates (two APXs 
and two SGRs) have been analyzed by \citet{ViKu} who found that the total 
energy content of the ejecta in these remnants does not appear to be different 
from the energy in remnants surrounding common NSs ($\approx 10^{51}$ erg). 
This result implies that either these NSs were not born rapidly rotating, 
thus challenging the $\alpha-\Omega$ dynamo scenario, or their initial spin 
energy must have been lost without appreciably energizing the 
surrounding ejecta.
\citet{DaSt} discussed the possibility that most of the initial spin energy 
of millisecond magnetars is realeased through
gravitational radiation. They concluded that, in order to account for the 
constraints derived by \citet{ViKu}, the internal toroidal field at 
birth should be in the $10^{16}$ G range, consistent with the inference by 
\citet{Ste05} for the field strength in SGR 1806-20 and, by extension, 
magnetars in general.
A similar conclusion about strong GW-driven spindown in newly formed magnetars 
was reached by \citet{Ar03}, based on a totally independent argument.\\
\citet{Buc07} have also studied the possibility that the huge spin energy of a 
millisecond magnetar is promptly released in the form of a highly collimated 
wind, if the 
external (dipole) field exceeds $10^{15}$ G. 
The wind breaks through the surrounding supernova ejecta and produces a 
GRB-like event without directly energizing the ejecta.
However, the currently estimated magnetar formation rate exceeds the rate of 
observed GRBs by 2 orders of magnitude. Hence, this process can involve at 
most a tiny fraction of newly formed magnetars \citep{Buc07}.\\
Thus, a number of astrophysical arguments point to possible field strengths in 
the $10^{16}$ G range in magnetar interiors and suggest a possibly 
relevant role of GW emission in the early evolution of these objects. The aim 
of this paper is to explore the physical consequences of this hypothesis and 
draw a state-of-the-art scenario, in light of all approximations and 
uncertainties. 
The paper is organised as follows:
in $\S$ \ref{earlyspindown} we introduce the model for the early spindown of a 
magnetar subject to both magnetodipole and GW torques and
study the main factors determining the GW emission efficiency of a newly 
formed magnetar. 
In $\S$  \ref{detect} we calculate the expected energy loss through GW
emission from newly born magnetars as a function of the relevant NS parameters 
and re-address the problem of signal detection already discussed by 
\citet{Ste05}. 
Finally, in $\S$ \ref{decayrate} we address the problem of the thermal and 
magnetic evolution of NSs, extending previous treatments of field decay to 
field strengths $\sim 10^{16}$ G. 
\section{Magnetar early spindown: electromagnetic vs. GW emission}
\label{earlyspindown}
A newly formed NS with a strong toroidal magnetic field has a prolate 
deformation induced by the field ($\epsilon_B$, defined in $\S$ 
\ref{ellipticity}) with symmetry axis along the magnetic axis. 
If the magnetic and spin axes are misaligned, with tilt angle $\chi$, the 
angular velocity vector ({\boldmath$\Omega$}) precesses (in the NS frame) 
around the fixed angular momentum vector ({\boldmath$L$}). 
\citet{CuJo01} (see also \citealt{Cut02}) studied the precession dynamics 
taking also into account the effects of the centrifugal deformation of the NS. 
They showed that 
the triaxial ellipsoid behaviour is, in this case, formally  equivalent to 
that of a biaxial rigid body,  as far as the precessional motion is concerned.
If $I_0$ represents the moment of inertia of the spherical NS, the eigenvalues
for the distorted NS are $I_1 =I_2 = I_0 -(1/3) \Delta I_{\Omega} - (1/3) 
\Delta I_B$ and $I_3  = I_1 + \Delta I_B$, where $\Delta I_{\Omega}$ 
represents the centrifugal deformation. Since the latter term affects all 
eigenvalues in the same way it does not enter explicitly in the precession 
dynamics, which are determined only by the magnetic deformation (see Appendix 
\ref{appendixa}). Hence, the focus here and in the following will be on the 
magnetic deformation only.

The spin energy of a rotating spheroid is minimized, at fixed angular momentum, 
when its moment of inertia is maximum. For a prolate figure, this is achieved 
when the symmetry axis is orthogonal to the spin axis. In the presence of 
internal dissipative processes, the magnetic axis of a prolate NS will thus be 
driven towards orthogonal rotation \citep{MeTa72, Jon76}, which maximizes the 
efficiency of GW emission\footnote{A dipole field induces an 
oblate deformation with the magnetic (symmetry) axis corresponding to the 
largest axis of inertia. In the presence of dissipation, the figure will thus 
be driven towards aligned rotation and no GW emission will result.} 
\citep{Cut02}.

For arbitrary $\chi$, GWs will be emitted at both the spin frequency and its
octave, with a total rate of energy emission (cfr. \citealt{CuJo01} and 
references therein)
\begin{equation}
\label{generalGW}
\dot{\mbox{\small{E}}}_{\mbox{\tiny{gw}}} = -
\frac{2}{5} \frac{G (\mbox{\small{I}} \epsilon_B)^2}{\mbox{\small{c}}^5}
~\omega^6~\mbox{\small{sin}}^2 \chi (1 + 15 \mbox{\small{sin}}^2 \chi)
\end{equation}
An orthogonal rotator emits GWs only at twice its spin frequency, at the rate 
given by eq. (\ref{generalGW}) with $\chi=\pi/2$
\begin{equation}
\label{gwemission}
\dot{\mbox{\small{E}}}_{\mbox{\tiny{gw}}} (\chi = \pi/2) = 
-~\frac{32}{5}\frac{G(\mbox{\small{I}}\epsilon_B)^2}{c^5}~\omega^6
\end{equation}
%
Given the generation mechanism for the superstrong internal field, the magnetic
axis is expected to be just slightly tilted, initially, to the spin axis and,
for small $\chi$, the GW luminosity is largely suppressed (eq. \ref{generalGW}).
Given the efficiency of magnetodipole radiation at birth, 
the orthogonalization process must thus be quick enough for strong GW 
emission to ensue promptly, in order to be competitive with magnetodipole 
radiation.

Strictly speaking, the spindown luminosity of a magnetic dipole rotating in 
vacuo depends only on the magnitude of the dipole component orthogonal to the 
spin axis. However, according to the standard pulsar model, 
an aligned rotator 
is expected to have a spindown luminosity comparable to that of an orthogonal 
rotator. Studies on the structure of force-free NS magnetospheres suggest 
that this is indeed the case, within a factor of order unity
\citep{CoKaFe99, Gru06, Spi06}. Based on these results, we assume a newly 
formed magnetar to have the magnetodipole luminosity of an orthogonal rotator. 
Further discussion of these issues is delayed to $\S$ \ref{evaluating}. 

We neglect here the effects of strongly magnetised winds from newly formed 
magnetars spinning at $\approx$ 1 ms period and restrict our attention to 
external dipole fields $\leq 5 \times 10^{14}$ G. Although such winds
can be extremely efficient in carrying away angular momentum (and spin 
energy) from the NS in just a few minutes,
their efficiency is expected to be negligible for dipole fields weaker than 
$(6\div 7) \times 10^{14}$ G \citep{Tho04, Buc06, Metz07}. \\
To summarize, we describe the early spin evolution of a newly born magnetar 
as being driven by both magnetodipole and GW torques. Hence, as the tilt 
angle $\chi$ increases, a sufficiently large (prolate) deformation and rapid 
initial rotation can make GW emission dominate angular momentum and rotational 
energy losses. 
For a given spindown torque, $\dot{\omega} = - \mbox{\small{K}}_{\alpha} 
\omega^{\alpha}$, the corresponding spindown timescale is 
$\tau_{\mbox{\tiny{sd}}} \equiv \omega/(2\dot{\omega})$. For convenience, in the
following we express the relative strength of the two torques at birth in 
terms of the ratio:
\begin{equation}
\label{timescales}
\frac {\tau_{\mbox{\tiny{d}},i}}{\tau_{\mbox{\tiny{gw}},i}} =   
\frac{\mbox{\small{K}}_{\mbox{\tiny{gw}}} \omega^2_i} 
{\mbox{\small{K}}_{\mbox{\tiny{d}}}} = \frac{\omega^2_i}{A} \equiv x
\end{equation}
where $A \equiv \mbox{\small{K}}_{\mbox{\tiny{d}}}/
\mbox{\small{K}}_{\mbox{\tiny{gw}}}$; GW losses are dominant for $x > 1$.

The complete spin-down equation we adopt is thus:
\begin{equation}
\label{pdot}
\dot{\omega} = -\frac{2}{3}\frac{\mu^2_d}{\mbox{\small{Ic}}^3}~\omega^3 
-\frac{\dot{\mbox{\small{E}}}_{\mbox{\tiny{gw}}}}{\mbox{\small{I}} 
\omega}~\omega^5 = -\mbox{\small{K}}_{\mbox{\tiny{d}}}~\omega^3-
\mbox{\small{K}}_{\mbox{\tiny{gw}}}~\omega^5
\end{equation}
where $\dot{\mbox{\small{E}}}_{\mbox{\tiny{gw}}}$ is given by 
(\ref{generalGW}), or by (\ref{gwemission}) when $\chi \approx \pi/2$.
%
\subsection{Damping of freebody precession and orthogonalization}
\label{wobbledamp}
We consider here in some detail the very first stages when the prolate NS is 
formed, its symmetry axis being just slightly tilted to the spin axis. We
focus in particular on those processes that might affect the orientation 
of the symmetry axis and thus promote (or prevent) prompt and strong GW 
emission.

As stated above, freebody precession of the prolate spheroid is eventually 
damped by internal viscous torques, which redistribute angular momentum (with 
no loss) inside the NS so as to minimize the spin energy. 
Following the analysis of \citet{Cut02}, \citet{Ste05} assumed that 
internal (viscous) dissipation of free precession occurrs on a (very short) 
timescale $\tau_{\mbox{\tiny{ort}}} \simeq 10^4$ P$_{\mbox{\tiny{prec}}}$ 
\citep{AlSa88}, where P$_{\mbox{\tiny{prec}}}$ ($< $ few seconds) is the free 
precession period. This estimate results from the crust-core coupling caused by 
the interaction between superfluid neutron vortices and relativistic electrons 
in the NS core, where only electrons follow the instantaneous rotation of the 
crust. The application of this prescription to a newborn NS is subject to 
important caveats: neutron pairing in a $^3P_2$ state in the core occurs at 
a temperature T$_{\mbox{\tiny{cond}}} < 2 \times 10^9$ K (\citealt{TD96} and
references therein, \citealt{Pag04}) and crust formation also 
occurs at a temperature $\sim$ a few $\times 10^9$ K. The coupling mechanism 
studied by \citet{AlSa88} does not apply when 
T $>$ T$_{\mbox{\tiny{cond}}}$, since there is no superfluid and, likely, not even
a proper crust. The NS is more like a self-gravitating, rapidly rotating, 
fully degenerate fluid mass. A proper account of its early cooling is thus 
of crucial importance in this context.

If direct Urca processes occurr in the densest parts of the core,
then cooling is extremely fast due to the very large neutrino luminosity. 
The NS temperature drops to $10^9$ K in a matter of minutes, as opposed 
to $\sim$ 1 yr in the case of modified Urca cooling (cfr. \citealt{Pag06} and 
references therein). Therefore, if newly born magnetars cool through direct 
Urca processes, crust formation and transition to superfluidity in the 
core occur very quickly. The crust-core coupling mechanism described by 
\citet{AlSa88} could thus operate soon after NS birth and lead to
the very short orthogonalization time estimated above.

In the opposite limit, when NS cooling is driven by only the modified Urca 
reactions, the evolution is more complex. 
The temperature in this case evolves as (\citealt{Ow98} 
and references therein, \citealt{Pag06})
\begin{equation}
\label{coolinghistory}
\frac{T(t)}{10^9~K} = \left[\frac{t}{\tau_c} + \left(\frac{10^9~K}{T_i}
\right)^6\right]^{-1/6}
\end{equation}
where $\tau_c \simeq$ 1 yr and $T_i \approx 10^{10}$ K is the initial 
temperature of the NS at the end of the $\alpha-\Omega$ dynamo phase 
\citep{TD96}. As we are interested in the cooling at 
T $>$ T$_{\mbox{\tiny{cond}}}$, no further neutrino-emitting reactions are 
considered, such as those occurring at T$\approx$ T$_{\mbox{\tiny{cond}}}$ in
the so-called 
``minimal cooling scenario'' (\citealt{Pag04} and references therein). 
%
%
According to eq. (\ref{coolinghistory}), the NS temperature will reach a 
value $ 2 \times 10^9$ K in about 5 days after formation. 
During this time, the crust-core coupling mechanism introduced by 
\citet{AlSa88} cannot operate. In the absence of other viscous processes, the 
prolate spheroid will thus not be orthogonalized and GW emission will remain 
highly suppressed during those early days. Therefore
magnetodipole radiation by a $10^{14}$ G external field will carry away most 
of the initial spin energy, thus spoiling the possibility of significant GW 
emission also at later times.

However, the prolate NS is subject to at least two 
further (and competing) mechanisms that can in principle alter the 
orientation of its symmetry axis. 
First, GWs will be emitted even for a small initial tilt angle $\chi_i$,
though at a small rate. 
The corresponding radiation reaction torque will cause the spin and magnetic 
axes to align, regardless of whether the spheroid is oblate or prolate 
\citep{CuJo01}. This process always acts so as to quench the GW emission 
efficiency.\\
On the other hand, freebody precession induces internal motions in the newly 
formed, fluid NS, that are required by the condition of hydrostatic 
equilibrium \citep{MeTa72, Jon76}.
Dissipation of these internal motions through bulk viscosity is potentially 
able to orthgonalize the symmetry axis of the spheroid (the magnetic field 
axis) relative to the angular momentum vector, thus increasing the efficiency 
of GW emission.\\
The relative strength of these two mechanisms will thus determine the early 
evolution of the angle $\chi$ and the possibility of newly formed magnetars to 
become efficient GW emitters. Below we study these two processes in more 
detail, showing that orthogonalization through bulk viscous damping is 
expected to always prevail on radiation reaction, in the parameter range of 
interest to our work.
\subsubsection{\textbf{Equation of state and related quantities}}
\label{eos}
For our aims, the equation of state (EOS throughout) of NS matter 
enters through its role in determining the global structural properties of the 
NS, such as the mass-radius relation and the moment of inertia. We dot not 
consider the detailed microphysics that determine the EOS; rather we adopt a
phenomenological approach, parametrizing all results in terms of the NS mass 
and radius. \\
As shown by \citet{LaPr01} most NS EOS's are well approximated by a polytrope 
of index $n=1$, with $P = k \rho^2$, as far as their global properties are 
concerned. Here $k$ is determined by the stellar radius (see Appendix 
\ref{appendixb}). We also adopt the approximate formula for the NS moment of 
inertia given by \citet{LaPr01}: I$ \approx 0.35 M R^2 \simeq 1.4 \times 10^{45} 
(\mbox{\small{M}}/ 1.4~\mbox{\small{M}}_{\odot}) (\mbox{\small{R}}/ 
12~\mbox{\small{km}})^2 $ g cm$^2$. Finally we note that, as discussed in 
\citet{LaPr07}, internal magnetic fields weaker than $\simeq 10^{18}$ G are 
not expected to sizeably affect the microphysics that determine the EOS. 

For the sake of completeness, we also report here the resulting relation 
between the measured $P$ and $\dot{P}$ of a NS and the corresponding magnetic 
field strength as derived by the magnetodipole spindown formula (see eq.
\ref{pdot}). 
%
\begin{eqnarray}
\label{dipolefield}
\mbox{\small{B}}_{\mbox{\tiny{d}}} & \simeq & 4.4\times10^{19}(\mbox{\small{P}}
\dot{\mbox{\small{P}}})^{1/2} \left(\frac{\mbox{\small{M }}}
{1.4 \mbox{\small{M}}_{\odot}}\right)^{1/2} \left( \frac{\mbox{\small{R }}}
{12 \mbox{\small{km}}}\right)^{-2}~\mbox{\small{G}} \nonumber \\
\mu_{\mbox{\tiny{d}}} & \simeq & 3.8\times10^{37}(\mbox{\small{P}} 
\dot{\mbox{\small{P}}})^{1/2} \left( \frac{\mbox{\small{M }}}
{1.4 \mbox{\small{M}}_{\odot}}\right)^{1/2} \left( \frac{\mbox{\small{R }}}
{12 \mbox{\small{km}}}\right)\mbox{\small{Gcm}}^3
\end{eqnarray}
\subsubsection{\textbf{The Magnetically-Induced Distortion of a Neutron Star}}
\label{ellipticity}
One of the key parameters of interest is the deviation from spherical symmetry 
($\epsilon_B$) of the NS, that is caused by the anisotropic pressure of the 
internal magnetic field. This determines the frequency of freebody precession 
and the GW luminosity of the rapidly spinning NS. 
In general, $\epsilon_B$ will be determined by the volume-integrated ratio of 
the magnetic to gravitational binding energy densities, times some numerical
coefficient accounting for field geometry and NS structure. Setting the 
latter factor to unity, and assuming an approximately constant density in the 
NS core
\begin{equation}
\label{epsilonb}
\frac{I_3 - I_1}{I_1}= \frac{\Delta I_B}{I_1}\equiv \epsilon_B = 
\frac{15}{4}\frac{E_{\mbox{\tiny{B}}}}{E_{\mbox{\tiny{G}}}} =
\frac{15}{4}\mbox{E}^{-1}_{\mbox{\tiny{G}}}\int \frac{B^2}{8\pi}~dV
\end{equation}
as determined from general arguments based on the virial theorem 
(\citealt{Cut02} and references therein). In the above definition, the 
integral is extended to the whole NS volume where the magnetic field is 
present. We are assuming the 
internal field to be mostly toroidal, the poloidal component being 
sufficiently small to be energetically negligible, as envisaged in the
millisecond magnetar formation scenario. E$_{\mbox{\tiny{G}}}$ in eq. 
(\ref{epsilonb}) is the gravitational binding energy of the NS that, for 
a polytrope of index $n=1$, is E$_{\mbox{\tiny{G}}} = (3/4) G M^2/R$. \\
The value of $\epsilon_B$ can thus be approximated as
\begin{equation}
\label{epsbvalue}
\epsilon_B \approx - 1.15 \times 10^{-3} \left(\frac{\mbox{\small{E}}_{
\mbox{\tiny{B}}}}{10^{50}~\mbox{\small{erg}}}\right) 
\left(\frac{\mbox{\small{R}}}{12\mbox{\small{ km}}}\right)
\left(\frac{\mbox{\small{M}}}{1.4\mbox{\small{M}}_{\odot}}\right)^{-2}
\end{equation}
that is negative for a prolate shape. The volume-averaged strength of the 
internal (toroidal) magnetic field is:
\begin{equation}
\label{energy-field}
\left< \frac{\mbox{\small{B}}_{\mbox{\tiny{t}}}}{2 \times 10^{16}~\mbox
{\small{G}}} \right> \simeq 0.93~\left(\frac{\mbox{\small{E}}_
{\mbox{\tiny{B}}}}{10^{50}~
\mbox{\small{erg}}}\right)^{1/2} \left(\frac{\mbox{\small{R}}}{12\mbox{
\small{km}}}\right)^{-3/2}
\end{equation}
Uncertainties in the exact value of 
$\epsilon_B$ as a function of the (unknown) internal field distribution can be 
parametrized by adding a further multiplicative factor\footnote{The exact 
value of $\epsilon_B$ is uncertain as it depends on the magnetic field 
configuration and EOS. Using a general relativistic treatment of the NS 
structure, \citet{BoGou} studied the way in which the numerical factor in 
eq. (\ref{epsilonb}) changes for different configurations of the internal 
magnetic field. These authors showed that this factor can be higher than 
15/4, or even much higher, for specific magnetic field geometries (for example
the case of an internal field that is mainly concentrated in an outer shell of 
the NS core, or the case of an intermittent field distribution rather than a 
uniform one). \citet{Has08} studied the same problem allowing for different 
EOS's and combinations of the toroidal and poloidal components of the internal 
field. Their results confirm that the magnetically-induced ellipticity can be 
in general somewhat larger, or significantly larger in particular cases, than 
the estimate given by eq. (\ref{epsilonb}) and (\ref{epsbvalue}). \\
Overall then, for a given magnetic energy, larger ellipticities than given by 
eq. (\ref{epsilonb}) are a possibility worth of further investigations.}
($\eta$) in the right-hand side of eq. (\ref{epsilonb}) where, in general, 
$\eta > 1$ is expected. We neglect this factor for clarity, but stress that 
the quantity E$_{\mbox{\tiny{B}}}$ used throughout can more generally be 
substituted with $\eta$E$_{\mbox{\tiny{B}}}$. We explicitly assume $\eta =1$ 
from here on.
\subsubsection{\textbf{Gravitational radiation reaction}}
\label{GRreaction}
The effect of gravitational radiation reaction on damping of freebody 
precession was investigated by \citet{CuJo01}, 
who showed that GW emission always drives the tilt angle $\chi$ to zero, 
independent of whether the NS is oblate or prolate. In the limit of small 
$\chi$, these authors express the freebody precession damping timescale 
($\tau_{\mbox{\tiny{rr}}}$) as: 
\begin{eqnarray}
\label{wobdamp}
\frac{\mbox{\small{sin}} \chi} {\mbox{\small{d(sin $\chi$)}} / 
\mbox{\small{dt}}} \equiv \tau^{\chi}_{\mbox{\tiny{rr}}}  =  
\frac{5 c^5}{2 G \Omega^4}\frac{I_1}{(I_1\epsilon_B)^2}  \approx & & \nonumber 
\\ 
~~~\approx 3.7 \left(\frac{\mbox{\small{E}}_{\mbox{\tiny{B}}}} 
{10^{50}\mbox{\small{erg}}}\right)^{-2} \left(\frac{\mbox{\small{P}}}
{\mbox{\small{ms}}}\right)^4\left(\frac{\mbox{\small{R}}}
{12\mbox{\small{km}}}\right)^{-4} & \mbox{d} & ~. 
\end{eqnarray}
The importance of this effect on the early evolution of a newly formed 
magnetar can be determined by comparing this timescale to the bulk 
viscosity dissipation timescale and to the magnetodipole spindown timescale.
\subsubsection{\textbf{Bulk viscosity in newly born NSs}}
\label{bulk}
As shown by \citet{MeTa72}, a field of internal motions is 
excited in order to maintain hydrostatic equilibrium everywhere within a 
fluid star undergoing freebody precession. 
These motions are periodic, with the same frequency as freebody precession and 
amplitude proportional to the (non-spherical) centrifugal deformation of the 
star ($\rho_{\Omega} \propto \Omega^2)$ and tilt angle $\chi$ of the 
magnetic axis. In this section we discuss the damping of such motions 
by bulk viscosity and the associated growth of the tilt angle $\chi$, within 
a non-superfluid NS made only of neutrons, protons and electrons 
($npe$ matter). 

The assumption of pure $npe$ matter corresponds to the least favourable case 
for damping through bulk viscosity and likely not the most realistic. 
At the highest densities of NS cores muons and possibly hyperons ($\Lambda, 
\Sigma^{-}$) are expected to appear, significantly increasing the bulk 
viscosity coefficient (e.g. \citealt{Jon76, LiOw02} and references therein). 
In this respect, our calculations are conservative. Detailed models should 
lead to significantly shorter damping times for the excited modes.\\
Fluid bulk viscosity is a consequence of the finite time it takes for the fluid 
to react to changes in one of its thermodynamic parameters, adjusting the 
others to their new equilibrium values. 
In hot NS matter, displacement of a fluid parcel from the equilibrium position 
causes departure from chemical equilibrium between particle species: bulk 
viscosity is thus determined by the activation of $\beta$-reactions trying to 
restore chemical equilibrium. A general expression for the bulk viscosity 
coefficient in NS matter ($\zeta$) has been derived by \citet{LiOw02}
\begin{equation}
\label{bulkgeneral}
\mbox{Re}(\zeta) = \frac{n \tau \left(\partial p/ \partial x\right)_n 
d\tilde{x} /dn} {1+ (\omega \tau)^2}
\simeq  \frac{n  \left(\partial p/ \partial x\right)_n d\tilde{x} /dn}
{\omega^2 \tau}~~,
\end{equation}
where $n$ is the particle density, $p$ is the pressure (whose derivative is 
calculated at constant proton fraction, $x = n_p /n_n \simeq n_p/n$) and 
 $\tilde{x}$ is the equilibrium value of $x$. The quantities in the 
denominator are the mode frequency ($\omega_{\mbox{\tiny{pre}}} \sim \epsilon_B 
\Omega_{\mbox{\tiny{spin}}}$) and $\beta$-reaction timescale ($\tau_{\beta}$). 
%
%
%
Defining the characteristic damping time of the excited motions as 
$
\tau_{\mbox{\tiny{d}}} \equiv 2 \mbox{E}_{\mbox{\tiny{pre}}}/
|\dot{\mbox{E}}_{\mbox{\tiny{pre}}}|
$
one can express it as (cf. \citealt{Ow98})
\begin{equation}
\label{deftau}
\tau_{\mbox{\tiny{d}}} = \left(\frac{\int{\zeta |\vec{\nabla} \cdot 
\vec{\delta v}|^2~ \mbox{\small{dV}}}}{2~\mbox{\small{E}}_{\mbox{\tiny{pre}}}}
\right)^{-1} = \frac{1}{\omega^2_{\mbox{\tiny{pre}}}} \left(\frac{\int{\zeta 
(\delta \rho / \rho)^2~ 
\mbox{\small{dV}}}}{2~\mbox{\small{E}}_{\mbox{\tiny{pre}}}}\right)^{-1}~,
\end{equation}
where $ \vec{\delta v}$ is the velocity perturbation associated to the mode. 
In the last step we have used the relation $\vec{\nabla} \cdot 
\vec{\delta v} = i \omega (\Delta n/n)$ \citep{LiOw02} and substituted the 
Eulerian perturbation, $\delta \rho$, to the Lagrangian one, $\Delta \rho$. 
A word of caution is needed here. The analysis by \citet{MeTa72} gives 
$\nabla \cdot \vec{\xi} = 0$, \textit{i.e} $\Delta \rho = 0$ and $\delta \rho 
= - \vec{\xi} \cdot \nabla \rho$, a condition that would imply the absence of 
bulk viscosity. However, this result is a direct consequence of having 
considered strictly adiabatic fluid motions\footnote{In fact, bulk viscosity
is a result of deviations from strict adiabaticity of perturbations.} 
within a radative (sub-adiabatic) layer, in a first-order perturbative
analysis.  
As opposed to this, a polytropic NS
would have an almost adiabatic gradient, in which case the calculations by
Mestel \& Takhar would leave $\nabla \cdot \vec{\xi}$ totally unconstrained,
revealing the need for a higher-order pertubation analysis. 
Indeed these authors discuss the (non)applicability of their conclusion to a 
convective (almost adiabatic) zone, at the end of their section 3. They 
discuss a number of approximations in their analysis that may well break in 
the case of a very rapid rotator (such as a millisecond magnetar) thus
introducing non-negligible higher-order terms in the perturbations. In these 
cases, the internal dynamics of the oblique rotator becomes much more 
complicated and a detailed analysis is beyond our scope here. 
 
In general, however, motions of the same order of magnitude as the one 
discussed by \citet{MeTa72} are always expected to occur, whose rate of 
expansion ($\nabla \cdot \vec{\xi}$) must be calculated explicitly. Here, 
in analogy to \citet{RG92} and to several treatments of bulk viscosity for 
$r$-modes \citep{LiOwMo98, Ow98}, we assume that $\Delta \rho$ would be of the 
same order of magnitude of $\delta \rho$, leaving more detailed analyses to 
future work. To stress the importance of this aspect we note that 
\citet{LiMeOw99} carried out a second-order analysis for $r$-mode damping 
through bulk viscosity and derived an order of magnitude longer dissipation 
timescale than those based on the assumption $\Delta \rho \sim 
\delta \rho$.   

We finally note that $\delta v \sim \epsilon_B \omega R \sim 10^7$ cm s$^{-1}$, 
while the Alfv\`{e}n velocity is $v_A = B/\sqrt{4 \pi \rho} \sim (10^8\div 
10^9)$ cm s$^{-1}$ for the parameters of interest. Alfv\`{e}n waves are thus 
very efficiently in maintining rigid rotation of the fluid star, despite the 
perturbation in principle introduced by $\vec{\xi}$.

The calculation of the perturbation amplitude ($\delta \rho / \rho$) 
is detailed and discussed in Appendix \ref{appendixb}. 
We report here our result for the damping timescale of freebody precession 
according to eq. (\ref{deftau})
%
\begin{eqnarray}
\label{taudamp}
\tau_{\mbox{\tiny{d}}}
  & \simeq & 13.5~\frac{\mbox{\small{cotan}}^2 \chi} {1+3\mbox{\small{cos}}^2 
\chi}~\left(\frac{\mbox{\small{E}}_{\mbox{\tiny{B}}}}{10^{50}
\mbox{\small{erg}}}\right) \left(\frac{\mbox{P}}{\mbox{\small{ms}}}\right)^2 
\nonumber \\
 & & \left(\frac{\mbox{M}}{1.4~\mbox{M}_{\odot}}\right)^{-1}
\left(\frac{\mbox{T}}{10^{10}\mbox{\small{K}}}\right)^{-6}~~\mbox{ s ,}
\end{eqnarray}
a fairly short time even for small
initial values of the tilt angle ($\chi_i$). 

Given $\tau_{\mbox{\tiny{d}}}$, we can eventually calculate the growth
time of the tilt angle, $\tau^{\chi}_{\mbox{\tiny{d}}}$. As shown in Appendix 
\ref{appendixa} damping of freebody precession reduces the NS spin energy,
at a constant angular momentum, by changing the tilt angle $\chi$. From
eq. \ref{A9} we get:
%
\begin{equation}
\label{equivalence}
\tau_{\mbox{\tiny{d}}} = \frac{\mbox{cos} \chi}{\dot{\chi}~\mbox{sin} \chi}
= \frac{\mbox{\small{cos}}^2 \chi} {\mbox{\small{sin}}
^2 \chi} \frac{\mbox{\small{sin}} \chi} {\mbox{\small{d(sin $\chi$)}} / 
\mbox{\small{dt}}} = \mbox{\small{cotan}}^2 \chi~\tau^{\chi}_{\mbox{\tiny{d}}}~.
\end{equation}
%
%
Here, $\tau^{\chi}_{\mbox{\tiny{d}}}$ is the timescale to be compared to 
$\tau^{\chi}_{\mbox{\tiny{rr}}}$ (eq. \ref{wobdamp}) in order to determine the 
relative importance of bulk viscosity and radiation reaction in the evolution 
of $\chi$. 
In the limit of small tilt angle, for which cos$\chi \simeq 1$, we have
\begin{equation}
\label{radreactobulkvisc}
\frac{\tau^{\chi}_{\mbox{\tiny{d}}}}{\tau^{\chi}_{\mbox{\tiny{rr}}}} \approx
10^{-5} \left(\frac{\mbox{\small{E}}_{\mbox{\tiny{B}}}}{ 
10^{50}~\mbox{\small{erg}}}\right)^3  \left(\frac{\mbox{\small{P}}}
{\mbox{\small{ms}}}\right)^{-2}  \left(\frac{\mbox{\small{T}}}
{10^{10}\mbox{\small{K}}}\right)^{-6} \left(\frac{\mbox{\small{R}}}{12
\mbox{\small{Km}}}\right)^4
\end{equation}
From this we derive the condition for bulk viscosity to largely prevail on 
gravitational radiation reaction, 
so that orthogonalization is essentially unaffected by radiation-reaction. 
For definiteness we require $\tau^{\chi}_{\mbox{\tiny{d}}} < 
0.1 \tau^{\chi}_{\mbox{\tiny{rr}}}$, a condition that translates to
\begin{equation}
\label{2radreactobulkvisc}
\mbox{\small{E}}_{\mbox{\tiny{B}}}  \leq  2.3 \times 10^{51}
\left(\frac{\mbox{\small{T}}}{10^{10}\mbox{\small{K}}}\right)^2 
\left(\frac{\mbox{\small{P}}}{\mbox{\small{ms}}}\right)^{2/3}  
\left(\frac{12\mbox{\small{km}}}{\mbox{\small{R}}}\right)^{4/3} 
\mbox{\small{erg}}
\end{equation}
or\footnote{For small $\chi$ the temperature is always very near to 
$10^{10}$ K. Therefore, this constraint is easily met if B is not very close
to 10$^{17}$ G.} B $< 9 \times 10^{16}$ (T/$10^{10}$K) 
(P$_{\mbox{\tiny{ms}}}$)$^{1/3}$ G.\\
%
Soon after formation, the temperature T decreases on a very short timescale,
not much different from $\tau^{\chi}_{\mbox{\tiny{d}}}$ itself, even for the slow
cooling given by eq. (\ref{coolinghistory}). The angle 
$\chi$ and T thus evolve on comparable timescales and just considering the 
damping timescale at a given temperature is not appropriate. Rather, the 
coupled evolution of $\chi$ and T must be solved in order to derive a 
reliable estimate of the time it takes for $\chi$ to grow to large values.

Inserting the cooling history (\ref{coolinghistory}) in (\ref{taudamp}) or
(\ref{equivalence}), the resulting equation for $\chi$ can be solved with 
initial conditions $\chi = \chi_i$ and T$_i = 10^{10}$ K. 
Since $\tau_{\mbox{\tiny{d}}}(t) = \mbox{N} ~T^{-6}_{10}(t)~
\mbox{\small{cos}}^2 \chi /[\mbox{\small{sin}}^2 \chi (1+3\mbox{\small{cos}}^2 
\chi)]$, the expression T$^{-6}_{10}$(t) = [(t/30) + 1] gives: 
\begin{equation}
\label{complete}
\frac{\mbox{cotan}\chi}{1+3\mbox{cos}^2 \chi} \mbox{d}\chi= 
\frac{\mbox{\small{dt}}}{\mbox{N} \left(\frac{t}{30} + 1 \right)}
\end{equation}
%
whose solution is:
\begin{eqnarray}
\label{chisolved}
\frac{\mbox{sin}^2 \chi}{1+3\mbox{cos}^2 \chi} & = &
\frac{\mbox{sin}^2 \chi_i}{1+3\mbox{cos}^2 \chi_i} \left(\frac{t}{30} +1
\right)^{240/\mbox{\small{N}}} = \nonumber \\
 & = & \frac{\mbox{sin}^2 \chi_i}{1+3\mbox{cos}^2 \chi_i}~
T_{10}^{-1440/\mbox{\small{N}}}
\end{eqnarray}
%
From the above we can obtain the time (in seconds) or the temperature (in units
of $10^{10}$ K) at which a sufficiently large value of the angle $\chi$ is 
reached starting from a given, small initial tilt angle $\chi_i$.\\
The requirement that the damping time of free precession be much shorter than
the initial spindown timescale of the NS allows two key constraints to be met 
jointly and our scenario to maintain full self-consistence; first, damping
of free precession will be well described in terms of the approximation of
constant angular momentum. Second, efficient GW emission will ensue quick 
enough for the NS \textit{initial} spin energy to be still fully available.

The initial timescale for electromagnetic spindown is 
$\tau^i_{\mbox{\tiny{em}}} \simeq 1.1~P^2_{\mbox{\tiny{ms}}} 
B^{-2}_{\mbox{\tiny{d}},14}$ d, and in a time $\simeq 0.1 \tau^i_{\mbox{\tiny{em}}}$
less than 10\% of the initial spin energy is lost to magnetic dipole 
radiation. Hence, we consider this as the longest time over which 
orthogonalization must take place for our scenario to apply.
Therefore, given $\chi_i =$ 1 deg, the angle $\chi$ will grow to a 
sufficiently large value - say, $\chi = 60$ deg\footnote{For $\chi \simeq 60$ 
deg GW emission is very efficient and can be approximated by eq. 
(\ref{gwemission})} - in a time shorter than $0.1~\tau^i_{\mbox{\tiny{em}}}$, 
if\footnote{Note that for any B$_{\mbox{\tiny{d}}} \geq 6 \times 10^{13}$ G 
and P$_{i,\mbox{\tiny{ms}}} \leq 3$ ms, 0.1$\tau_{i,\mbox{\tiny{em}}} \leq $ 3 days. 
The core temperature will accordingly be $ > 2 \times 10^9$ K, so that
our condition covers essentially the range of temperature before the core
becomes superfluid.} 
%
%
%
\begin{eqnarray}
\label{constraint}
\frac{\mbox{\small{E}}_{\mbox{\tiny{B}}}}{10^{50}~\mbox{\small{erg}}} & < & 2.1
~\frac{\mbox{\small{M}}_{1.4}}{\mbox{\small{P}}^2_{\mbox{\tiny{ms}}}} 
\left[ \mbox{\small{ln}}\left(320~\mbox{\small{M}}_{1.4}
\mbox{\small{R}}^{-4}_{12} \frac{\mbox{\small{P}}^2_{\mbox{\tiny{ms}}}}
{\mbox{\small{B}}^2_{\mbox{\tiny{d,14}}}} + 1 \right) \right] \simeq \nonumber \\
 & & 4.2~\frac{\mbox{\small{M}}_{1.4}}{\mbox{\small{P}}^2_{\mbox{\tiny{ms}}}} 
\left(3 + \mbox{\small{ln}} \frac{\mbox{\small{P}}_{\mbox{\tiny{ms}}}}
{\mbox{\small{B}}_{\mbox{\tiny{d,14}}}} +
\mbox{\small{ln}} \frac{\mbox{\small{M}}_{1.4}}{\mbox{\small{R}}^4_{12}} 
\right)
\end{eqnarray}
Note that the numerical coefficient in the last step is $\sim$ 5 if $\chi_i =$ 
2 deg, and $\sim$ 2.8 if $\chi_i =$ 0.1 deg.

\begin{figure}
\begin{center}
\leavevmode\includegraphics[width=2.25in, height=3.11in, angle = 
-90]{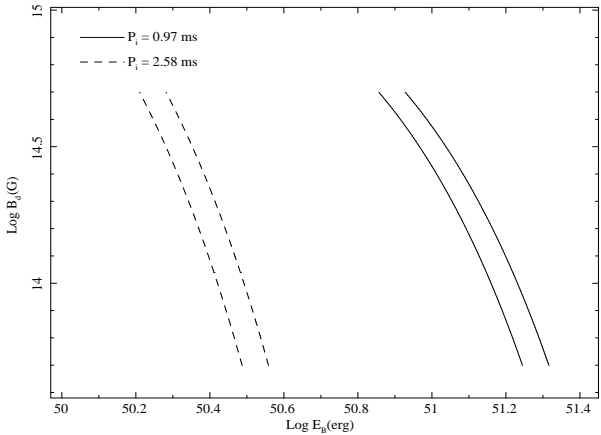}
\caption{The region of parameter space, in the external dipole field strength 
(B$_{\mbox{\tiny{d}}}$) vs. internal magnetic energy (E$_{\mbox{\tiny{B}}}$), where 
the timescale for the angle $\chi$ to grow to a large value (60 deg) is less 
than one tenth of the initial spindown timescale through magnetodipole 
radiation. The favourable region lies on the left of the corresponding 
limiting curve. For each value of the initial spin period, two curves are 
plotted for two different values of the initial tilt angle, $\chi_i =1$ (left) 
and 2 deg (right). Note that further decreasing $\chi_i$ from 1 deg 
to 0.1 deg shifts the curves to the left by a slightly larger amount than 
the decrease of $\chi_i$ from 2 deg to 1 deg (cfr. eq. \ref{constraint}). 
 This would still leave ample room in parameter space for the fastest spinning 
magnetars, while it would drastically reduce the available parameter space 
for magnetars with P$_i \geq$ 2.5 ms.}
\label{figconstraint}
\end{center}
\end{figure}

Eq. (\ref{constraint}) is represented in Fig. \ref{figconstraint} by two
curves corresponding to initial spin periods of 0.97 and 2.58 ms, that 
approximately bracket the relevant range of spin periods for newly formed
magnetars. 
%
Each curve is plotted for two different values of the initial tilt
angle, $\chi_i= 1$ and $2$ deg (see caption for further details).

The constraint shown in Fig. \ref{figconstraint} will be used, together with 
independent constraints derived in later sections, to identify the region in 
parameter space where GW emission efficiency from newly formed magnetars 
is optimized. 
We stress again here that all our calculations were carried out by assuming 
pure $npe$ matter, a conservative assumption that minimises the efficiency 
of bulk viscosity. 
Finally, in Appendix \ref{appendixb} we calculate the centrifugal distortion 
of the density profile in the slow-rotation limit, which is also likely to 
underestimate the deformation of a millisecond magnetar. This translates into 
an underestimate of $\delta \rho$ in eq. (\ref{deftau}) and, thus, of the 
efficiency of bulk viscosity. 

We conclude that, despite having chosen a ``worst case approximation'',  
damping of freebody precession through
bulk viscosity in a newly formed, rapidly spinning and strongly magnetized NS 
can be very efficient in the parameter range considered here. 
Therefore, strong GW emission from an almost orthogonal, rapidly rotating NS 
is likely to ensue quickly as a consequence of a strong toroidal magnetic 
field \textit{and} of the efficient dissipation of its freebody precession 
energy. The implications of this are explored further in the next section.
\section{Amplitude and detectability of the emitted signal}
\label{detect}
\citet{Ste05} calculated the expected signal-to-noise ratio (S/N) 
that a putative GW signal from a newly born magnetar would have, for an 
optimal (matched-filter) detection, adopting the broadband design 
sensitivity of Advanced Ligo. At frequencies between 0.5 and 2 kHz this is 
well approximated by
\begin{equation}
\mbox{\small{S}}_h(f) \approx \mbox{\small{S}}_0~f^2~~,
\end{equation}
where $\mbox{\small{S}}_h(f)$ represents the one-sided spectral noise 
distribution of the detector, $f = 2 \omega$ the GW signal frequency and 
S$_0 \simeq 2.1 \times 10^{-53}$ Hz$^{-1}$ (\citealt{OwLi02, Cut02} and 
references therein). Note that the designed sensitivity for Advanced Virgo is 
very similar, in this range of frequencies \citep{Los07}, so that our 
calculations hold essentially for both detectors.
frequency, $f = \omega/\pi$ for an orthogonal prolate rotator.\\
We re-address here this point to better qualify the role of the NS ellipticity 
in the detectability of the signal. We also correct a (small) numerical error 
in the calculated S/N curves in Fig. 1 of \citet{Ste05}, whose conclusions 
maintain their general validity.
\begin{figure*}
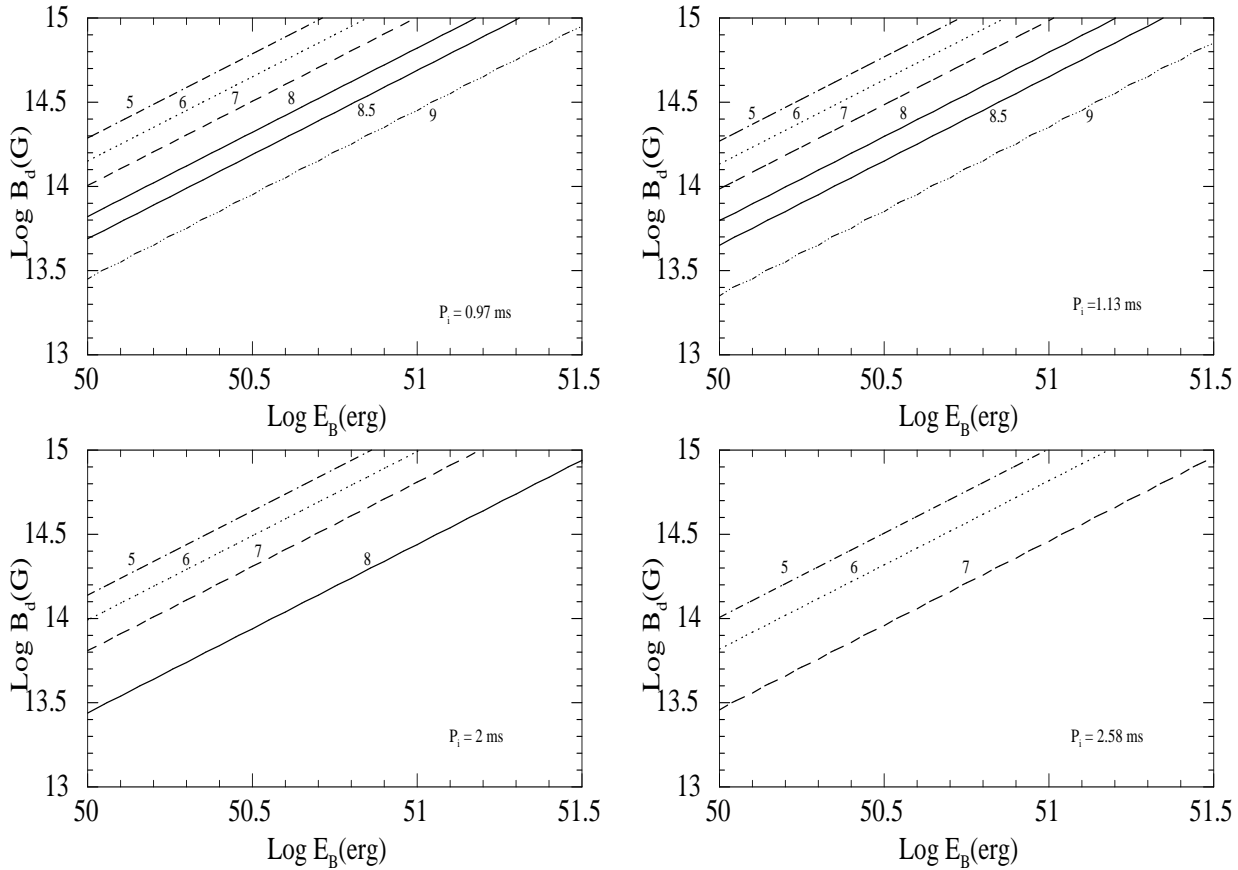

\begin{center}
\leavevmode\includegraphics[width=2.25in, height=3.11in, angle = 
-90]{SN097.ps}
\hspace{3.mm}
\includegraphics[width=2.25in, height=3.11in, angle = -90]{SN113.ps}
\hspace{3.mm}
\includegraphics[width=2.25in, height=3.11in, angle = -90]{SN2.ps}
\hspace{3.mm}
\includegraphics[width=2.25in, height=3.11in, angle = -90]{SN258.ps}
\caption{The optimal (matched-filtered) signal-to-noise for searches of the
GW signal from newly formed magnetars in the Virgo Cluster with Advanced Ligo
/VIrgo, calculated through eq. (\ref{integralnumber}). Four different values 
of the initial spin (indicated in the figures) are considered and results are 
shown as contour levels (at the values of S/N indicated on each line) in the 
B$_{\mbox{\tiny{d}}}$ vs. E$_{\mbox{\tiny{B}}}$ plane.}
\label{figSN}
\end{center}
\end{figure*}
The S/N of an optimal signal search is defined as 
%
\begin{equation}
\label{StoN}
\mbox{\small{S/N}} = 2 \left[ \int \frac{|\tilde{h}(f)|^2}{S_h(f)} 
\right]^{1/2}~.
\end{equation}
Here $\tilde{h}(f)$ is the Fourier transform of the instantaneous signal 
strain $h[f(t)]$ that, in the stationary phase approximation, is expressed as 
(cfr. \citealt{OwLi02} and references therein)
\begin{equation}
\label{stationaryphase}
|\tilde{h}(f)|^2 = \frac{1}{2} h^2[f(t)] \left| \frac{\mbox{\small{d}}f}
{\mbox{\small{dt}}}\right|^{-1}~~, 
\end{equation}
where the time derivative of $f$ is obtained from eq.(\ref{pdot}), since 
$\dot{f} = \dot{\omega}/ \pi$.
We adopt the expression for the strain amplitude - averaged over source 
orientation - given by\footnote{Comparing with the ``optimally-oriented'' 
strain amplitude $h_0$ given by \citet{Abb07}, we obtain the relation 
$h_a = 4/(5 \sqrt{3\pi}) h_0$.} \citet{UsCuBi00}:
\begin{equation}
\label{strainamplitude}
h_a (f) = \frac{16}{5} \left(\frac{\pi^3}{3}\right)^{1/2} 
\frac{{\mbox{\small{GI}} \epsilon_B}}{\mbox{\small{Dc}}^4}~f^2~~,
\end{equation}
where D is the source distance. Further averaging over the detector antenna 
pattern,
eq. (\ref{StoN}) gives the optimal S/N ratio:
\begin{eqnarray}
\label{integral}
 & & \mbox{\small{S/N}} = \sqrt{\frac{2}{5}} \left[ \int \frac{ h^2(t)}{S_h(f) 
\mbox{ \small{d}}f/\mbox{\small{dt}}} \right]^{1/2}  = \nonumber \\
& & = \frac{4}{5} \sqrt{\frac{\pi \mbox{\small{GI}}}{6 \mbox{\small{c}}^3}} 
\frac{\pi}{\mbox{\small{DS}}^{1/2}_0} 
\left(\frac{\mbox{\small{K}}_{\mbox{\tiny{gw}}}}{\mbox{\small{K}}_{\mbox{\tiny{d}}}}
\right)^{1/2}\left[2~\mbox{\small{ln}}\frac{f_i}{f_f} - \mbox{\small{ln}}
\frac{a + f^2_i}{a+ f^2_f}\right]^{1/2} 
\end{eqnarray}
where we have set $a = \mbox{\small{K}}_{\mbox{\tiny{d}}}/ (\pi^2 
\mbox{\small{K}}_{\mbox{\tiny{gw}}}) = A/ \pi^2$. Substituing the numerical
values:
\begin{eqnarray}
\label{integralnumber}
\mbox{\small{S/N}} \approx & 6 \left(\frac{\mbox{\small{E}}_{\mbox{\tiny{B}}}}
{10^{50}\mbox{\small{erg}}}\right) \left(\frac{\mbox{\small{B}}_{\mbox{\tiny{d}}}}
{10^{14}\mbox{\small{G}}}\right)^{-1} \left(\frac{\mbox{\small{R}}}
{12\mbox{\small{km}}}\right) \left(\frac{\mbox{\small{M}}}
{1.4\mbox{\small{M}}_{\odot}}\right)^{-1/2} & \nonumber  \\
& \left(\frac{\mbox{\small{D}}}
{20 \mbox{\small{Mpc}}}\right)^{-1} \left [ 2 \mbox{{\small{ln}}} 
\frac{f_i}{f_f} + \mbox{\small{ln}} \frac{a + f^2_i}{a + f^2_f}\right]^{1/2}
&
\end{eqnarray} 
%
%
In Fig. \ref{figSN} we show the curves of constant S/N in the B$_{\mbox{
\tiny{d}}}$ vs. E$_{\mbox{\tiny{B}}}$ for selected values of the 
initial spin, as derived from eq. (\ref{integralnumber}). More details are
given in the caption.
According to eq. (\ref{integral}) the \textit{maximum} S/N is obtained in the 
limit $a \rightarrow 0$, which depends \textit{only} on the initial spin energy 
of the NS \textit{and not} on its oblateness. Clearly, this value 
$[\mbox{\small{S/N}}]_{\mbox{\tiny{MAX}}}$ is attained as the magnetodipole 
spin-down torque disappears, so that all of the initial spin energy of the NS 
is lost to GWs. 
This can also be seen directly by substituting the expression for the 
pure GW-driven spindown in eq. (\ref{integral}), which gives
%
%
\begin{eqnarray}
\label{onlyGW}
\left[\mbox{\small{S/N}}\right]_{\mbox{\tiny{MAX}}} & \simeq & 4.5 
\left(\frac{\mbox{\small{D}}}{20\mbox{\small
{Mpc}}}\right)^{-1} \left(\frac{\mbox{\small{R}}}{12 \mbox{\small
{Km}}}\right) \left(\frac{\mbox{\small{M}}}{1.4\mbox{\small
{M}}_{\odot}}\right)^{1/2} \nonumber \\
& & \left[\left(\frac{f_f}{\mbox{\small{kHz}}}\right)^{-2} - 
\left(\frac{f_i}{\mbox{\small{kHz}}}\right)^{-2} \right]^{1/2} 
\end{eqnarray}
\subsection{Quantifying the gravitational and electromagnetic energy output}
\label{comparetoSNRs}
Thus far we have discussed the conditions under which the hypothesis that 
newly born magnetars be detectable sources of GWs with next generation 
detectors can be true. X-ray observations of Supernova Remnants (SNRs) around 
magnetar candidates in the Galaxy give us clues on the actual viability of 
this hypothesis (cf. $\S$ \ref{intro}). In this section, we show that our 
scenario is indeed fairly consistent with such observations. We find that, 
if magnetars at birth were \textit{detectable} sources of GWs from 
Virgo-cluster distance, then SNRs around them would likely show no significant
excess of energy injection with respect to ``ordinary'' 
SNRs\footnote{Assuming GWs do not appreciably transfer energy to the Supernova 
ejecta}

We begin calculating the total integrated energy emitted via GWs by a 
newly born magnetar as:
\begin{equation}
\label{integrategw}
\Delta \mbox{\small{E}}_{\mbox{\tiny{gw}}} = - \int_{t_i}^{\infty} 
\dot{\mbox{\small{E}}}_{\mbox{\tiny{gw}}}~dt = 
\int_{0}^{\omega_i} \frac{\dot{\mbox{\small{E}}}_{\mbox{\tiny{gw}}}}
{\dot{\omega}}~d\omega
\end{equation}
%
For the spindown model in eq. (\ref{pdot}), the above integral gives
\begin{eqnarray}
\label{solveintegral}
\Delta \mbox{\small{E}}_{\mbox{\tiny{gw}}} & =  I \int_{0}^{\omega_i} 
\frac{\omega^3}{\omega^2 + A}~d\omega = I\left[\frac{\omega^2}{2} - 
\frac{A}{2}\mbox{ln}(\omega^2+A) \right]^{\omega_i}_0 = & \nonumber \\
& = \mbox{\small{E}}_{\mbox{\tiny{spin}},i} \left[1 - \frac{A}{\omega^2_i}
\mbox{\small{ln}} \left(1+ \frac{\omega^2_i}{A}\right) \right] &
\end{eqnarray}
%
Eq. (\ref{solveintegral}) expresses $\Delta \mbox{\small{E}}_{\mbox{\tiny{gw}}}$ 
as a fraction ($\delta$) of the initial spin energy of the NS. 
The remaining energy $\Delta \mbox{E}_{\mbox{\tiny{em}}} = \mbox{\small{E}}_{\mbox{\tiny{spin}},i} (1-\delta)$ is the maximum that can be
transferred to the SN ejecta. 
Next, we define the quantity $\Delta \mbox{\small{E}}_{\mbox{\tiny{tr}}}
= \beta \Delta \mbox{\small{E}}_{\mbox{\tiny{em}}}$ as the amount of that 
energy that is effectively transferred to the ejecta. In the absence of 
additional competing torques and/or a physical prescription for the transfer 
mechanism, we will assume it to be perfectly efficient ($\beta=1$), although 
leaving the explicit dependence on $\beta$ in the calculations that follow. \\
Recalling the definition of $x$ as the ratio of gravitational to magnetic 
dipole torque \textit{at birth} (eq. \ref{timescales})
%
%
%
and using eq. (\ref{solveintegral}) we then get
%
%
%
\begin{equation}
\label{secondrelation}
\frac{\mbox{\small{ln}}(1+x)}{x} = 
\frac{\Delta \mbox{\small{E}}_{\mbox{\tiny{em}}}}
{\mbox{\small{E}}_{\mbox{\tiny{spin}},i}} = \frac{1}{\beta}
 \frac{\Delta \mbox{\small{E}}_{\mbox{\tiny{tr}}}}
{\mbox{\small{E}}_{\mbox{\tiny{spin}},i}} \leq  
\frac{1}{\beta} \frac{E^{\mbox{\tiny{inj}}}_{SNR}}
{\mbox{\small{E}}_{\mbox{\tiny{spin}},i}}
\end{equation}
where $E^{\mbox{\tiny{inj}}}_{SNR}$ is the amount of energy transferred to the 
ejecta as determined from observations \citep{ViKu}. Observations provide an 
upper bound to this quantity, from which the inequality in the last step in 
eq. (\ref{secondrelation}) obtains. 
We derive constraints on magnetar parameters at birth by using the upper bound 
determined by \citet{ViKu} discussed in this paper and our GW emission scenario.

First fix a value for $x$: eq. (\ref{secondrelation}) gives the maximum 
value of the initial spin energy or, equivalently, 
($\omega^2_i$)$_{\mbox{\tiny{max}}}$ compatible with the upper limit on 
$\mbox{\small{E}}^{\mbox{\tiny{inj}}}_{\mbox{\tiny{SNR}}}$. Accordingly, the maximum 
value of $A = (\omega^2_i)_{\mbox{\tiny{max}}}/x$ is constrained (cfr. eq. 
\ref{timescales}):
\begin{equation}
\label{amax}
A_{\mbox{\tiny{max}}} =  \frac{(\omega^2_i)_{\mbox{\tiny{max}}}}{x} =
\frac{2 \mbox{\small{E}}^{\mbox{\tiny{inj}}}_{\mbox{\tiny{SNR}}}}
{\beta \mbox{\small{I }} \mbox{\small{ln}}(1+x)}
\end{equation}
\begin{figure*}
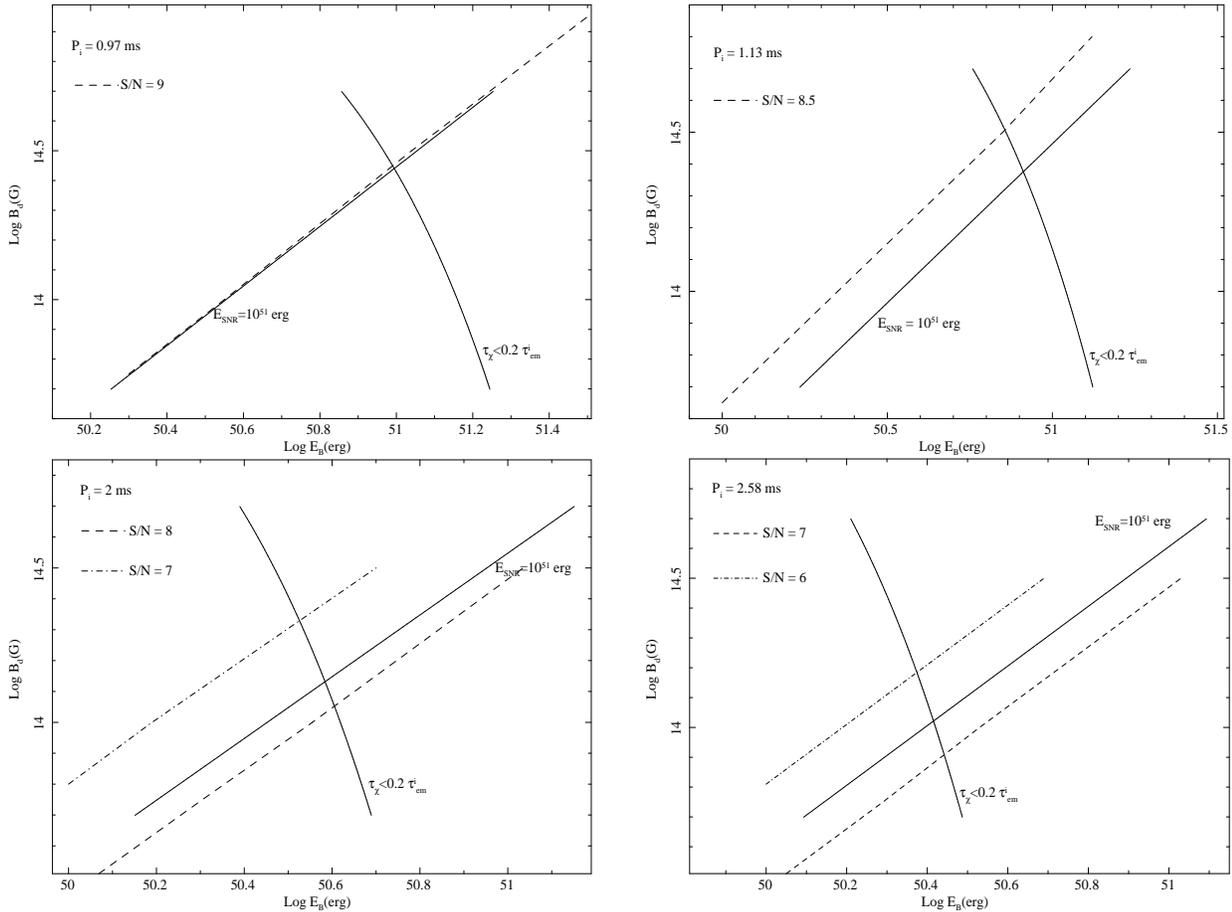

\begin{center}
\leavevmode\includegraphics[width=2.37in, height=3.06in, angle = 
-90]{graphriassunt097.ps}
\hspace{5mm}
\includegraphics[width=2.37in, height=3.06in, angle = -90]{graphriassunt113.ps}
\hspace{5mm}
\includegraphics[width=2.37in, height=3.06in, angle = -90]{graphriassunt2.ps}
\hspace{5mm}
\includegraphics[width=2.37in, height=3.06in, angle = -90]{graphriassunt258.ps}
\caption{Curves defining the region of parameter space - in the 
B$_{\mbox{\tiny{d}}}$ vs. E$_{\mbox{\tiny{B}}}$ plane -  where all constraints for 
efficient emission and detection of GWs are fulfilled, for four initial spin 
periods. The region on the left of (and below) the solid curve labelled 
$\tau_{\chi} < 0.1 \tau^i_{\mbox{\tiny{em}}}$ is where the tilt angle $\chi$ 
becomes $\geq 60$ deg in less than one fifth of the initial magnetodipole 
spindown time. The region on the right of (and below) the straight, 
solid curve is where the energy emitted electromagnetically by the newly 
formed magnetar is $\leq 10^{51}$ erg. It is obtained from eq. 
(\ref{conclusive}) with four different values of $x=150, 100, 22, 10$ from top 
to bottom and right to left, corresponding to the initial spins indicated in 
the figures. The region on the right of the dashed (or dot-dashed) curves is 
where a match-filtered search with Advanced LIGO/Virgo would give a 
signal-to-noise ratio greater than indicated in the figures, with the source 
at the distance of the Virgo Cluster (20 Mpc).}
\label{graficofinale}
\end{center}
\end{figure*}
%
Inserting eq. (\ref{amax}) in the definition of A= K$_{\mbox{\tiny{d}}}/$ 
K$_{\mbox{\tiny{gw}}}$ (cfr. $\S$ \ref{earlyspindown}) we get the following 
condition, for the electromagnetic output of a newly born, millisecond 
spinning magnetar to be always $\leq 10^{51}$ erg: 
\begin{eqnarray}
\label{conclusive}
\mbox{\small{E}}_{\mbox{\tiny{B}}} \geq \sqrt{\frac{\mbox{\small{G}}}
{117.6 ~A_{\mbox{\tiny{max}}}}} \mbox{\small{Mc}}\mbox{\small{B}}_{\mbox{\tiny{d}}} =
 \sqrt{\frac{\beta \mbox{\small{G ln}}(1+x)}{672 \mbox{\small{ E}}
^{\mbox{\tiny{inj}}}_{\mbox{\tiny{SNR}}}}} 
\mbox{\small{cM}}^{3/2}\mbox{\small{R}}\mbox{\small{B}}_{\mbox{\tiny{d}}} 
\approx & & \nonumber \\
\approx 1.6 \times 10^{50}\sqrt{\beta \mbox{\small{ln}}(1+x)} 
\left(\frac{\mbox{\small{B}}_{\mbox{\tiny{d}}}}{10^{14}\mbox{\small{G}}}
\right) 
\left(\frac{\mbox{\small{M }}}{1.4\mbox{\small{ M}}_{\odot}}\right)^{3/2}
 & & \nonumber \\ 
\left(\frac{\mbox{\small{R}}}{12 \mbox{km}}\right)
\left(\frac{\mbox{\small{E}}^{\mbox{\tiny{inj}}}_{\tiny{SNR}}}{10^{51} 
\mbox{\small{erg}}}\right)^{-1/2}  &  &
\end{eqnarray}
Therefore, given $x$ one obtains $\omega_i$ and, eventually, the appropriate 
curve in the B$_{\mbox{\tiny{d}}}$ vs. E$_{\mbox{\tiny{B}}}$ plane. 
Results are shown in Fig. \ref{graficofinale}, where curves in the $B_
{\mbox{\tiny{d}}}$ vs.E$_{\mbox{\tiny{B}}}$ plane have been drawn for four 
different values of $x = 10, 22, 100$ and 150, corresponding to the initial
spins indicated in each plot. Curves of given signal-to-noise ratio from Fig. 
(\ref{figSN}) are also plotted together with curves defining the region where 
the orthogonalization timescale is sufficiently fast (see Fig. 
\ref{figconstraint}). 

We conclude that there exists a wide region of the parameter space where all 
constraints for efficient GW emission from newly born, millisecond spinning 
magnetars are met jointly. 
%
%
\section{Electromagnetic spin-down at birth}
\label{evaluating}
%
The ideal magnetodipole spindown formula (cfr. eq. 
\ref{pdot}) implies that a sizeable fraction 
of the initial spin energy of a millisecond spinning 
magnetar would be lost to electromagnetic emission, for B$_{\mbox{\tiny{d}}} 
\geq 3\times 10^{14}$ G. 

SGRs, and two out of seven AXPs, have spin periods and period derivatives 
leading to estimated dipole fields of a few times $10^{14}$ G, up to 
$\sim 10^{15}$ (cfr. \citealt{WoTh06} and references therein), some of which 
are apparently out of the optimal range for GW emission at birth, if they
maintained their dipole field since then.  
Although relation (\ref{pdot}) is certainly true to a good degree of 
approximation, the ideal magnetodipole formula does not appear to hold 
exactly in the few isolated NSs for which sufficiently accurate timing 
measurements exist. 

Observations show that the spin-down of pulsars, assumed to be wholly due 
to an electromagnetic torque, might be a somewhat weaker function of the spin 
frequency than expected in the ideal case. 
Indeed, in the few pulsars where the second time derivative of the spin 
frequency ($\ddot{\nu}$) could be measured, a ``braking index'' $n = 
(\ddot{\nu} \nu/\dot{\nu}^2) <3$ is typically derived. In particular, the Crab 
pulsar has a measured $n=2.5$ \citep{LyPrSm88} and 
PSR J1119-6127 has $n=2.9$ \citep{Cam00}. \citet{Liv07} 
have recently measured braking indices of three relatively young X-ray pulsars. 
Again, all results give $n<3$ (2.14 for PSR B0540-69, 2.84 for PSR B1509-58 
and 2.65 for PSR J1846-0258). The Vela pulsar has the largest known deviation 
from the ideal case, with $n =1.4$ \citep{Lyn96}, although 
subtraction of the frequent glitches from its spin history presents
difficulties and that result should be taken with some caution.\\
Similar (or even greater) uncertainties are associated to the putative 
$n\simeq 6$ determined by \citet{Mar04} for PSR J0537-6910 
(the ``Big Glitcher''). This source is found to exhibit very frequent 
glitches; post-glitch ``recoveries'' are known to significantly affect the 
second time derivative of the spin frequency. 
For this reason, the measured values of $\ddot{\nu}$ from such a frequent 
glitcher are probably unsuitable to determine reliably the
``true'' secular spindown trend. 
\citet{Mid06} discuss in detail several aspects concerning the 
timing properties of this source, arguing for a much smaller value of $n$.

%
%
The examples of $10^3 \div 10^4$ yrs old pulsars with braking indices 
lower than 3 suggest considering the same possibility for magnetar candidates,
which indeed have similar ages and similarly high glitch activity \citep{Dal03, 
Isr07, Dib08}. No measurements of $n$ has yet been obtained for AXPs and SGRs. 
It would be interesting to know from the data whether these sources 
share this same spindown property of other isolated NSs. 

Several mechanisms able to produce a braking index smaller than 3 have been
proposed in the literature, since the early measurements of $n$ in the Crab 
pulsar. \citet{BlaRo88} discuss in general the effects of evolving 
the parameters of the magnetodipole torque. They identify a secular increase 
in the surface magnetic field, at least over the first $\sim 10^3 \div 10^4$
years, as a most promising explanation for $n<3$ in young NSs. 
Amplification of crustal fields through a thermomagnetic battery over this 
early phase had been previously discussed (\citealt{BlApHe83} and references 
therein), which could find application also in this context. \\
Alternatively, several suggestions have been made for a secular increase of 
the tilt angle between the magnetic dipole and the spin axis, based on a wide 
variety of physical mechanisms \citep{Gol70, Rud91, LiEpBa92, RuZhCh98}. 
More recently, the possibility that the zone of closed field lines in the 
magnetosphere (the corotating region) does not reach the light cylinder has
been widely discussed, both in the general context of studies of pulsar 
magnetospheres \citep{CoKaFe99, Gru06, CoSp06, Spi06} and in relation to the 
magnetospheric structure of newly born magnetars \citep{Buc06, Metz07}.
 
Simulations by \citet{Spi06} have shown that the closed field line region 
reaches the light cylinder in a matter of one (at most) rotation period, as a 
model NS is set into rotation. However, it is not obvious whether the region 
of closed field-lines can subsequently track the expansion of the light 
cylinder as the NS spins down or, rather, 
lag behind it at an increasing distance, which would naturally lead to $n<3$
(\citealt{CoSp06} and references therein).

Without indicating any particular mechanism, we discuss the consequences that 
an electromagnetic braking index $n<3$ would have, at a purely phenomenological 
level. \\
%
%
Considering a generic model for electromagnetic spindown with braking index 
$n$, one has
\begin{equation}
\label{relatethem}
\dot{\omega}^{(n)} = -\mbox{\small{K}}^{(n)}_{\mbox{\tiny{d}}}~\omega^n~~, 
\end{equation}
where the constant $\mbox{\small{K}}^{(n)}_{\mbox{\tiny{d}}}= 
(\dot{\omega}_0/\omega^n_0)$ and  the subscript ``0'' refers to present-day 
values of the parameters.
According to eq. (\ref{relatethem}), given measured timing parameters 
($\omega_0, \dot{\omega}_0$) of a source the expression for its spindown at 
birth in the case $n\neq$ 3 and case $n=3$ are related through
\begin{equation}
\label{lessthan3}
\dot{\omega}^{(n)}_i =  -\mbox{\small{K}}^{(n)}_{\mbox{\tiny{d}}} 
\omega^n_i 
= - \mbox{\small{K}}^{(3)}_{\mbox{\tiny{d}}} \omega^3_0
\left(\frac{\omega_i}{\omega_0}\right)^n = \dot{\omega}^{(3)}_i
\left(\frac{\omega_i}{\omega_0}\right)^{n-3}~~,
\end{equation}
%
%
where the subscript ``$i$'' indicates quantities at birth. \\
Note that the quantity in parentheses is smaller than unity for $n<3$. It is
therefore natural to ask what value of $n$ would be required in magnetar 
candidates (AXPs/SGRs), given their measured $\omega_0$ and $\dot{\omega}_0$,  
for their spindown 
\textit{at birth} to have been dominated by
GW emission rather than by the magnetic dipole radiation.

In order to answer this, we define the maximum allowed strength of the 
electromagnetic spindown at birth ($\dot{\omega}_{i,\mbox{\tiny{max}}}$). Given
the results of the previous section, this maximum value will equal that the 
ideal magnetodipole formula ($n=$3) would give for B$_{\mbox{\tiny{d}}} = 
\mbox{\small{B}}_{\mbox{\tiny{d}}} (\mbox{\small{max}}) = 2 \times 10^{14}$ G
(a value that guarantees strong GW emission at birth, cfr. $\S$ 
\ref{comparetoSNRs}). 
Therefore we write, $\dot{\omega}_{i,\mbox{\tiny{max}}} = 
\mbox{\small{K}}^{(3)}_{\mbox{\tiny{d}},\mbox{\tiny{max}}}~\omega^3_i$ and the last 
step of eq. (\ref{lessthan3}) must be smaller than 
$\dot{\omega}_{i, \mbox{\tiny{max}}}$, which gives:
\begin{equation}
\label{uppern1}
\mbox{\small{K}}^{(3)}_{\mbox{\tiny{d}}} \left(\frac{\omega_i}{\omega_0}
\right)^{n-3} \leq \mbox{\small{K}}^{(3)}_{\mbox{\tiny{d}}}(\mbox{\small{max}})
\end{equation}
From this, since the ratio of the torque functions corresponds to the ratio
of the magnetic dipole fields:
\begin{eqnarray}
\label{uppern}
(3-n) \mbox{\small{Log}}\left(\frac{\omega_i}{\omega_0}\right) & \geq &
2~\mbox{\small{Log}}\left( \frac{\mbox{\small{B}}_{\mbox{\tiny{d}}}}
{\mbox{\small{B}}_{\mbox{\tiny{d}}}~(\mbox{\small{max}})}\right)~~~~~~~~~~~
\mbox{\small{or}} \nonumber \\
n & \leq & 3 - 2~\frac{\mbox{\small{Log}}\left[\mbox{\small{B}}_{\mbox{\tiny{d}}}
/ \mbox{\small{B}}_{\mbox{\tiny{d}}}
(\mbox{\small{max}})\right]}{3+\mbox{\small{Log}} \mbox{\small{P}}_0 - 
\mbox{\small{Log}} \mbox{\small{P}}_{i,\mbox{\tiny{ms}}}}
\end{eqnarray}
Among SGRs and AXPs, SGR 1806-20 has the strongest inferred dipole field 
($\simeq 1.1 \times 10^{15}$ G with R$= 12$ km and M$= 1.4$M$_{\odot}$, cfr. 
\citealt{WoTh06}), thus requiring the largest deviation from $n=3$ in the 
hypothesis discussed here. Even assuming a (relatively) slow initial spin 
period for this source, P$_i = 3$ ms, eq. (\ref{uppern}) gives $n\leq 2.6$, 
a wholly plausible value compared to other isolated NSs.
Note that the constraint is slightly weaker for other SGRs and/or considering a 
spin period at birth shorter than 3 ms. For AXPs, the limit on $n$ ranges from
2.7 to 2.85, for an (unfavourable) initial spin of 3 ms.
This speculative argument would require direct measurements of braking 
indices in magnetar  candidates. However, our aim here was to emphasize the 
dependence (strong, in some cases) of the calculations of previous sections 
on a number of poorly constrained physical parameteres, and the importance of 
further studies on all of the above aspects.
%
%
\section{The decay of core fields in the $10^{16}$ G range}
\label{decayrate}
The secular evolution (and dissipation) of the magnetic field in NS cores was 
studied in detail by Goldreich \& Reisenegger (1992) (\citealt{GR92} from here 
on) and their analysis was extended by \citet{TD96}  
to the specific case of magnetar fields in the $10^{15}$ G range. In 
magnetars, large-scale field instabilities leading to fast dissipation events 
were studied in detail as well \citep{TD95, TD01, Lyu03}, in order to interpret
the powerful bursts and flares of SGRs. 

In GR92 three separate processes for secular field evolution were identified 
two of which, ohmic dissipation and ambipolar diffusion, are dissipative while 
the third, Hall drift, conserves magnetic energy. 
In particular, ambipolar diffusion was found to be more sensitive to the 
field intensity (GR92) which in fact implies that, while this process is not 
very important in normal NSs, it is the main mechanism for direct field 
decay in magnetars (TD96). \\
Hall drift can affect indirectly the evolution and dissipation of 
magnetic fields in NS interiors, on longer timescales than 
those characteristic of ambipolar diffusion \citep{TD96, Arr04}. As 
suggested in GR92 (and recently studied in detail by \citealt{Cu04}), 
excited Hall modes of field diffusion could decay (or cascade) to shorter 
wavelenghts, that are subject to enhanced ohmic dissipation. This has 
particular relevance for accelerating field decay in a magnetar's crust, as 
recent studies pointed out \citep{Pon07, PoGe07}. 
Furthermore, Hall diffusion can drive - yet conserving the total energy - an 
initially stable MHD configuration close to a new equilibrium configuration
 with smaller total energy. A point can be reached where the field suddenly 
relaxes to the new equilibrium, if (sufficiently fast) fluid motions are 
allowed within the stably stratified NS interior. 

As long as the early (ages much less than $\sim 10^4$ yr) evolution of 
magnetars is concerned, however, ambipolar diffusion in the NS core is 
expected to be the dominant mode of field decay.\\
Ambipolar diffusion drives a slow motion of charged particles with respect to 
background neutrons, which is opposed by both particle friction and chemical 
potential gradients in the stably stratified NS medium. \citealt{GR92} 
identified two 
separate modes of ambipolar diffusion, differing by their effect on chemical 
composition. The solenoidal mode does not perturb chemical equilibrium and 
thus is counteracted only by particle friction. The irrotational mode, on the 
other hand, does perturb chemical equilibrium and cannot evolve on timescales 
shorter than the $\beta$-reaction timescale. \\
As shown in \citealt{TD96}, $\beta$-reactions are very efficient at 
erasing chemical equilibrium imbalance when $T>T_{\mbox{\tiny{tr}}} \approx
5.73 \times 10^8~(\rho_{15}/0.7)^{1/12}$ K.
At these high temperatures, both modes of ambipolar diffusion are effectively 
opposed by neutron-proton friction only. Field decay occurs on the same 
timescale in both modes (GR92)
\begin{eqnarray}
\label{GR92}
t^{(early)}_d  = \frac{4 \pi n^2_e}{\lambda B^2} \left(\frac{L}{a}\right)^2 
\simeq ~~~~~~~~~~~~~~~~~~~~~~~~~~~~~~~~~~~~~&  &  
\nonumber \\
~~~~~~~\simeq  2.2\times 10^4\left(\frac{T}{10^9\mbox{\small{K}}}\right)^2
\left(\frac{\rho_{15}}{0.7}\right)^{\frac{2}{3}}\left(\frac{B}{10^{16}\mbox
{\small{G}}}\right)^{-2}~\mbox{\small{yr}} & &,
\end{eqnarray}
where 
$L$ and $a$ are the characteristic scale of variation of the Lorentz 
force and chemical potential, respectively,  
$(L/a) \approx 9.16~T^4_9\rho^{-1/3}_{15}(L/2\mbox{ km})$ (GR92). The latter is 
$\gg 1$ at high temperatures, given the high efficiency of $\beta$-reactions, 
while it becomes $<1$ as the temperature drops
and the efficiency of $\beta$-reactions decreases. At T$ > 10^9$ K, the 
timescale (\ref{GR92}) is much longer than the NS age or its cooling 
timescale. Therefore, field decay is negligible as long as the
temperature is this high.

As NS cooling proceeds, the point is reached (at $T\leq T_{\mbox{\tiny{tr}}}$) 
where chemical equilibrium imbalance becomes the main obstacle against which 
magnetic stresses must work to drive particle diffusion. This affects 
only the irrotational mode, while the solenoidal mode still decays on the 
timescale of eq. (\ref{GR92}). Hence, the two modes grow at different rates 
with the irrotational mode evolving on a longer timescale\footnote{The two 
timescales are formally equal at $T=T_{\mbox{\tiny{tr}}}$, but the irrotational 
mode becomes quickly much slower below $T=T_{\mbox{\tiny{tr}}}$. For example, a 
20\% decrease of $T$ below $T=T_{\mbox{\tiny{tr}}}$ gives a 6 times longer decay 
timescale for the irrotational mode.} (TD96, GR92)
\begin{eqnarray}
\label{tirr}
t^{(late)}_d = \frac{4\pi n^2_e}{\lambda B^2} = \left(\frac{L}{a}\right)^{-2} 
t^{(early)}_d \approx ~~~~~~~~~~~~~~~~~~~~~~~~~~&  & \nonumber \\
\approx 7\times 10^3 \left(\frac{T}{T_{\mbox{\tiny{tr}}}}\right)^{-6} 
\left(\frac{\rho_{15}}{0.7}\right)^{5/6}\left(\frac{L}{2\mbox{\small{km}}}
\right)^{\frac{3}{2}} \left(\frac{B}{10^{16}\mbox{\small{G}}}\right)^{-2}
\mbox{\small{yr}} & &.  
\end{eqnarray}
%
TD96 considered  
in detail this lower-temperature regime. Based on stability arguments, these 
authors suggested that the solenoidal mode is expected to carry just a small 
fraction of the magnetic energy, most of it being tapped by the irrotational 
mode. The main conclusion of this scenario is that only a tiny fraction of the 
magnetic energy reservoir in the NS core is lost either in the high-T regime,
or via field decay through the solenoidal mode at lower temperatures. Most of 
the magnetic energy dissipation occurs via the slow decay of the irrotational 
mode, at $T \leq T_{\mbox{\tiny{tr}}}$.
This is based on the assumption that field decay, and the irrotational mode in
particular, is effectively frozen at $T>T_{\mbox{\tiny{tr}}}$. 

The above scenario holds for fields in the $10^{15}$ G range. According to 
eq. (eq. \ref{GR92}), field decay at higher temperatures 
is significantly faster for stronger magnetic fields. Further, 
dissipation of even a small fraction of the magnetic energy reservoir may in 
principle affect NS cooling, if the reservoir is sufficiently large. In 
fact, field decay at T$> 10^9$ K is not frozen if B is larger than $10^{16}$ G. 
\begin{figure}
\begin{center}
\includegraphics[width=2.6in, height=3.35in, angle=-90]{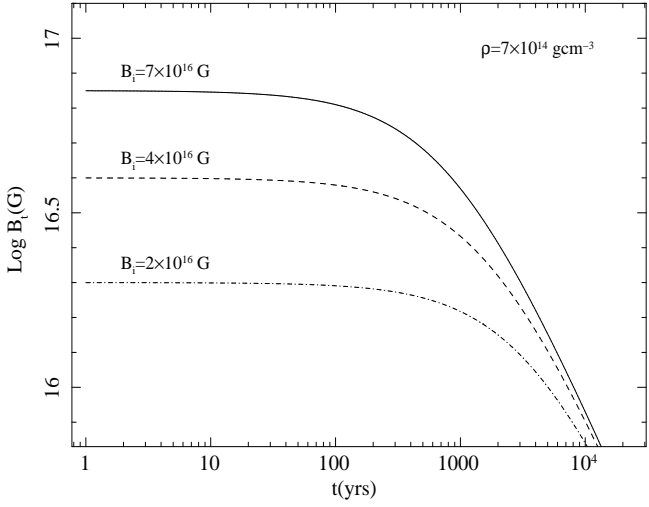}
\caption{Expected evolution of the magnetic field intensity - at a specific 
value of the density $\rho = 7 \times 10^{14}$ g cm$^{-3} = 
2.5 \rho_{\mbox{\tiny{nuc}}}$ - for three selected values of the initial 
intensity (B$_{\mbox{\tiny{i}}}$). Temperature equilibrium according to eq. 
(\ref{Tevolving}) is assumed at each time. The tracks converge, as the field 
strength approaches $\approx 7\times 10^{15}$ G, towards the asymptotic 
solution given by \citet{TD96}.}
\label{Bevolve}
\end{center}
\end{figure}
In analogy to the treatment of TD96, we check here whether an equilibrium 
condition between heating and cooling in the high-$T$ regime can apply as well, 
with field decay described by eq. (\ref{GR92}). For uniformity with that work
we adopt the same normalizations for the parameters used by \citet{TD96}.

We can write the heating rate per unit volume through field decay 
%
\begin{equation}
\label{heatirrot}
\frac{dU^+}{dt} = \frac{B^2}{4\pi t^{(early)}_d} \approx 3.69 \times 10^{19}
~\frac{B^4_{16}}{T^2_9~\rho^{\frac{2}{3}}_{15}}~~
\mbox{erg cm}^{-3}\mbox{s}^{-1}
\end{equation}
while the cooling rate per unit volume through modified Urca reactions is
\begin{equation}
\label{coolURCA}
\frac{dU^-}{dt} \simeq 9.6 \times 10^{20}~T^8_9~\rho^{\frac{2}{3}}_{15}~
\mbox{erg cm}^{-3}\mbox{s}^{-1}~~.
\end{equation}
Equating the two rates gives the equilibrium temperature
\begin{equation}
\label{equilibriumT}
T_{\mbox{\tiny{eq}}} \simeq 6.6 \times 10^8\left(\frac{B}{10^{16}~
\mbox{\small{G}}}\right)^{\frac{2}{5}} \left(\frac{\rho_{15}}{0.7}\right)^
{-\frac{2}{15}} \left(\frac{L}{2\mbox{Km}}\right)^{-\frac{1}{5}}~~\mbox{
\small{K}}
\end{equation}
Note that T$_{\mbox{\tiny{eq}}}$ is higher than $T_{\mbox{\tiny{tr}}}$ if $B\geq 
7 \times 10^{15}~(\rho_{15}/0.7)^{13/24}(L/2\mbox{km})^{-1/8}$ G. Fields 
larger than that would thus be able to
dissipate enough energy and balance neutrino cooling even in the early phase 
when the solenoidal and irrotational mode are still degenerate. 
This conclusion describes a regime that was not considered in Thompson \& 
Duncan (1996): very strong magnetic fields ($\sim 10^{16}$ G) decaying and 
heating a NS core at very high temperatures ($\sim 10^9$ K) and at very young
ages (years to centuries). The resulting evolution of magnetic field and 
temperature are coupled, as already shown by TD96. Our 
solution (eq. \ref{equilibriumT}) joins smoothly the one found by 
\citealt{TD96} (their eq. 31) in the sense that both give the same value of 
the magnetic field strength ($B\approx 7 \times 10^{15}$ G) when calculated at 
$T_{\mbox{\tiny{tr}}}$.
The two regimes are, in this sense, complementary, forming a continuous 
evolutionary sequence through  T$_{\mbox{\tiny{tr}}}$ for an arbitrarily large 
magnetic field whose decay is driven by ambipolar diffusion.\\
In order to better illustrate this, we calculate here the joint evolution of
the magnetic field strength and the equilibrium temperature, according to the
equilibrium conditions discussed above. Consider the rate of magnetic energy 
dissipation per unit volume
\begin{equation}
\label{evolveT}
\frac{B}{4\pi}\frac{dB}{dt} = - \frac{B^2}{4\pi t^{(early)}_d}~~,
\end{equation}
Inserting eq. (\ref{GR92}) and (\ref{equilibriumT}) in (\ref{evolveT}) gives
\begin{equation}
\label{evolveB}
\frac{dB_{16}}{dt} \approx - 3.12\times 10^{-12} \left(\frac{\rho_{15}}{0.7}
\right)^{-\frac{2}{5}}~B^{\frac{11}{5}}_{16} \left(\frac{L}{2\mbox{km}}\right)
^{-\frac{8}{5}}~~,
\end{equation}
whose solution is
\begin{equation}
\label{Bevolving}
B_{16}(t) = \left[1.12 \times 10^{-4} \left(\frac{\rho_{15}}{0.7}\right)^
{-\frac{2}{5}}~\frac{t}{\mbox{yr}} + \left(\frac{1}{B_{i,16}}\right)^
{\frac{6}{5}}\right]^{-\frac{5}{6}}
\end{equation}
with $B_i$ the strength of the magnetic field at the initial time $t_i$.
As an illustrative example, in Fig. \ref{Bevolve} we show the evolution of the 
core magnetic field according to eq. (\ref{Bevolving}), for three different 
initial values.
The corresponding equilibrium temperature evolution is:
\begin{equation}
\label{Tevolving}
T_{8,eq} \simeq 6.6 \left[1.1 \times 10^{-4}\frac{t}{\mbox{\small{
yr}}}\left(\frac{\rho_{15}}{0.7}\right)^{-\frac{2}{5}} +\left(\frac{1}
{B_{i,16}}\right)^{\frac{6}{5}}\right]^{-\frac{1}{3}}
\end{equation}
We caution that the equilibrium temperature as a function of density is in
principle different from the actual temperature profile throughout the NS core. 
Finding this would require solving the heat flux problem self-consistently, 
which includes account for the strong suppression of heat transport across 
magnetic field lines in a superstrong magnetic field.

\begin{figure}
\begin{center}
\includegraphics[width=2.6in, height=3.35in, angle=-90]{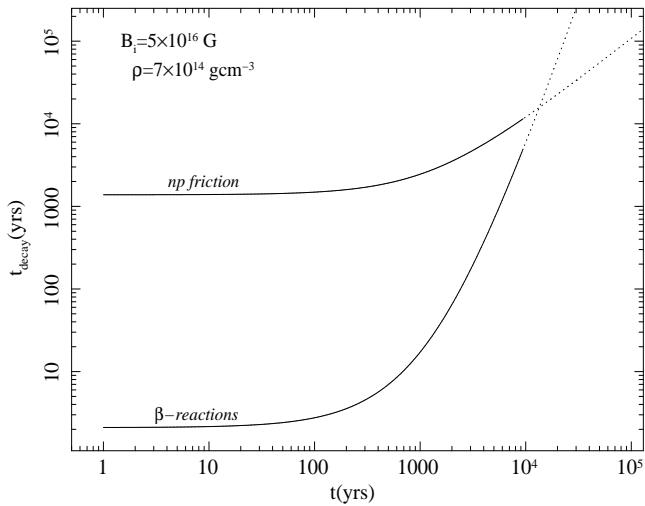}
\caption{Evolution of the ambipolar diffusion decay timescales determined by 
each of the two friction mechanisms, separately, calculated at a particular 
density $\rho = 2.5 \rho_{\mbox{\tiny{nuc}}}$ and for a given initial field 
strength B$_{\mbox{\tiny{i}}} = 5\times 10^{16}$ G. The timescales evolve 
according to their dependence on the temperature T and magnetic field B. 
The evolution of T(t) and B(t) is given by eqs. (\ref{Tevolving}) and 
(\ref{Bevolving}) and is thus determined by $np$-friction alone, initially. 
However, the $\beta$-reaction timescale is a much stronger function of time 
than the $np$-friction timescale, in this regime. Curves are dotted where 
the two timescales become comparable, within a factor of a few, and the 
approximation under which timescales are calculated - through eqs. 
(\ref{Bevolving}) and (\ref{Tevolving}) - does not hold anymore. At this 
point, the temperature is close to the transition temperature 
T$_{\mbox{\tiny{tr}}}$ and the magnetic field close to the value $\approx 7 
\times 10^{15}$ G (see text).}
\label{comparetimes}
\end{center}
\end{figure}

With the above expressions for the equilibrium regime, the decay timescales of 
the two modes of ambipolar diffusion can be evaluated self-consistently - 
given an initial magnetic core field B$_{\mbox{\tiny{t}},i}$ - and their 
values compared. 
In Fig. \ref{comparetimes} we show, for illustration, the evolution of the 
two timescales for an initial (uniform) core magnetic field 
B$_{\mbox{\tiny{t}},i} = 5\times 10^{16}$ G and at a given value of the density 
$\rho_{15} = 0.7$. Clearly, field evolution is always determined by the 
longest decay time: as long as particle friction dominates, both modes decay 
on the timescale $t^{(early)}_d$. Once chemical equilibrium imbalance overtakes 
particle friction (at $t\approx 10^4$ yrs in our example), the two modes 
split: the solenoidal mode is unaffected and continues to evolve on the (now 
shorter) time $t^{(early)}_d$, while the irrotational mode is now subject to a 
slower evolution, determined by the efficiency of $\beta$-reactions. \\
We show in Fig. \ref{Bevolve} the evolution of the core magnetic field in the 
regime described here, for three different initial values. More details are 
given in the captions.

We can also estimate the time $\Delta t$ after which the temperature, once 
equilibrium between heating and cooling holds, reaches the transition 
value\footnote{This time is zero by definition for $B_i= 7 \times 10^{15}$ G} 
$T_{\mbox{\tiny{tr}}}$
\begin{eqnarray}
\label{deltatime}
\frac{\Delta t}{\mbox{yr}} & \approx & 1.36 \times 10^4\left(\frac{\rho_{15}}
{0.7}\right)^{-\frac{1}{4}}\left(\frac{L}{2\mbox{km}}\right)^{\frac{7}{4}} 
 \times \nonumber \\
& \times & \left[1 - 0.657 \left(\frac{1}{B_{i,16}}\right)^{\frac{6}{5}}
\left(\frac{\rho_{15}}{0.7}\right)^{-\frac{8}{5}}\left(\frac{L}{2\mbox{km}}
\right)^{\frac{32}{35}}\right] 
\end{eqnarray}
This gives $\Delta t \sim$ 4500 yrs for $B_i = 10^{16}$ G and has an 
asymptotic value $(\Delta t)_{\mbox{\tiny{max}}} \simeq 1.36 \times 10^4$ yrs for 
very large $B$. For $B_i=3\times 10^{16}$ G, $\Delta t \simeq 1.1 \times 10^4$ 
yrs, $\sim$ 80\% of the asymptotic value. 
The comparison with a NS standard cooling scenario, without any heat sources, 
is striking, since in this case $T_{\mbox{\tiny{tr}}}$ is reached in somewhat 
less than 10 yrs. \\
Once the field has decayed to the limiting value $\simeq 7 \times 10^{15}$ G,
the temperature equals $T_{\mbox{\tiny{tr}}}$ and the results obtained by
\citet{TD96} hold.
We conclude that the early thermal evolution of the core can be significantly 
altered (slowed down) by ultrastrong field decay. 
A detailed analysis of the consequences of this conclusion is beyond our scope
here and will be the subject of future study.
\subsection{Comparison with recent studies of core heating in magnetars}
\label{comparison}
Surface temperatures inferred from the X-ray spectra of SGRs/AXPs ($> 0.35$ 
keV, or $> 4 \times 10^6$ K) are a factor $\sim 2\div 3$ higher than expected 
from cooling calculations of NSs with their estimated ages, $\sim 10^4$ yrs
(\citealt{Kam07}). 
The heat source internal to the NS needs thus not only be strong enough to 
provide the excess energy but it also needs to be efficiently converted to 
\textit{surface} heating.

\citet{Kam07} considered how the surface temperature of a cooling NS would 
be affected when a (slowly decaying) heating source in its interior were 
included, in order to account for the heating provided by the decaying 
magnetic field. 
These authors suggest that, if heat is released in the core or deep 
crustal layers, only a minor fraction of it is transported to the surface and 
radiated thereby as X-ray photons.  Rahter, most of the released heat produces 
an enhancement in neutrino emission - which is strongly 
temperature dependent - and is thus radiated locally (via neutrinos). 
In fact, in their calculations the local temperature in the core is affected 
only slightly by internal heating. Core temperatures never reach $10^9$ K 
and the surface can't ever be kept as warm as AXPs/SGRs surface emission 
indicates.
\citet{Kam07} conclude that heat is most likely released in the outer 
regions of the crust, where neutrino-emitting processes are much less
efficient and heat is thus most efficiently transported to the surface, with 
only minor losses. 

Their results are therefore at variance with ours, that envisage an efficient 
heating of the core up to very high temperatures (as a function of the 
magnetic field strength).
\begin{figure}
\begin{center}
\includegraphics[width=2.6in, height=3.35in, angle=-90]{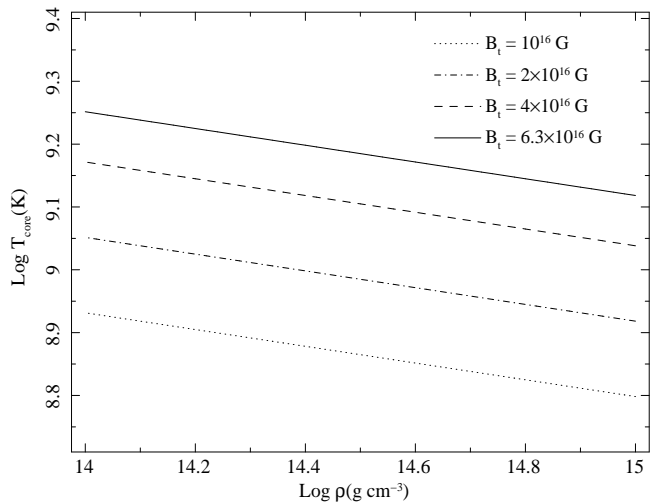}
\caption{Value of the equilibrium temperature (eq. \ref{Tevolving}) as a 
function of density throughout the NS core, for four selected values of the 
magnetic field strength, as indicated in the figure.}
\label{coretemp}
\end{center}
\end{figure}
The difference bewteen the two conclusions is readily found in the 
parametrization of the heating source chosen by \citet{Kam07}, which
corresponds to a different physical scenario.

Their model heat source is independent on the magnetic field strength and on 
the NS internal temperature. The maximum initial heating rate they consider 
(Q$_0 = 3 \times 10^{20}$ erg/s) effectively corresponds to our expression 
(\ref{heatirrot}) with B$\simeq 1.6 \times 10^{16}$ G. The heating rate in 
their model decreases exponentially in time, with a fixed time constant 
comparable to, but somewhat longer than, the source estimated age.\\
As opposed to this, we have $i)$ allowed for initially stronger fields and, 
thus, larger heating rates and $ii)$ considered a time-varying decay 
timescale, since the ambipolar diffusion-driven field decay is a function 
of the NS internal temperature and field strength (eq. \ref{heatirrot}).
In particular, the values of T and B at each epoch are determined 
self-consistently by the equilibrium condition between cooling and heating.
 
As a consequence, in the model by \citet{Kam07} the released heat does not 
affect the value of Q while it enhances neutrino emission. The latter being
a strong function of the temperature, the net result is a slight enhancement 
of the interior temperature, to make neutrinos able to carry away almost all 
the eccess heat. 

%
In ambipolar-diffusion driven field decay, on the other hand, heating has a
feedback on both cooling \textit{and} heating itself. As long as the 
temperature is very high the NS cools, with heating providing just a minor 
perturbation to the dominant process of $\nu$-cooling.
As the core temperature drops, however, the field decay rate grows (eq. 
\ref{heatirrot}) while neutrino emission drops. Eventually, the two rates 
become almost equal and at this stage their equilibration plays a key role, 
as stressed by \citet{TD96}. 
\begin{figure}
\begin{center}
\includegraphics[width=2.75in, height=3.35in, angle =-90]{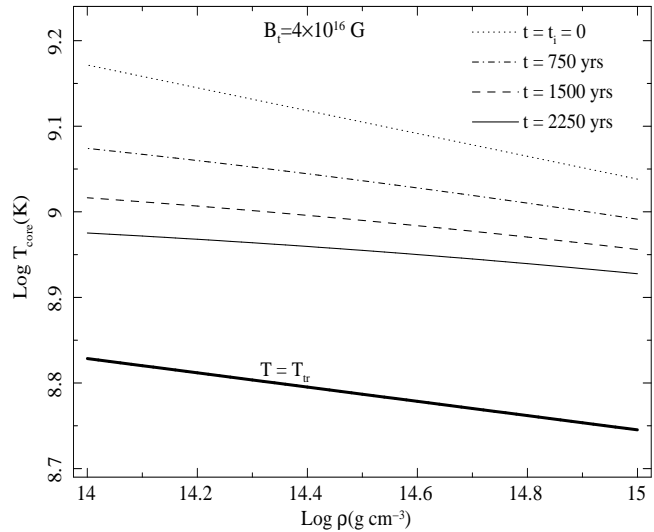}
\caption{The evolution of the temperature profile within a magnetar core 
($10^{14}$ g cm$^{-3} < \rho \leq 10^{15}$ g cm$^{-3}$) for a specific 
initial value of the toroidal field strength (B$_{\mbox{\tiny{t}}} = 
4\times 10^{16}$ G), corresponding to a total magnetic energy 
E$_{\mbox{\tiny{B}}} \simeq 4.6 \times 10^{50}$ erg or an ellipticity 
$\epsilon_{\mbox{\tiny{B}}} \approx - 4.6 \times 10^{-3}$.
The four curves describe the temperature profile at four different epochs, the
thick solid curve at the bottom defines the temperature at (and below) which
chemical equilibrium imbalance becomes the major limiting factor for ambipolar
diffusion of the irrotational mode (the regime described in TD96). Above the
thick line, the irrotational and solenoidal mode are degenerate and the 
regime described in the previous section holds. }
\label{corEvolve}
\end{center}
\end{figure}
Near equilibrium,
temperature variations have a strong feedback on both heating and cooling - 
and with opposite effects. This forces the temperature toward 
T$_{\mbox{\tiny{eq}}}$ (eq. \ref{equilibriumT}): 
as the field dissipates the temperature drops slighlty and a new equilibrium 
between heating and cooling is reached, at a slightly smaller 
temperature and field strength. 
The equilibrium temperature within the NS core, as a function of density, 
obtained through our eq. (\ref{equilibriumT}) is shown in Fig. 
(\ref{coretemp}) for four different values of the average core magnetic field
strength. In Fig. (\ref{corEvolve}) we show, for a given initial magnetic field 
strength, the equilibrium temperature throughout the magnetar core at four 
different epochs of its evolution according to eq. (\ref{evolveT}).
More details are given in the caption.\\
Therefore, the internal heating source considered by \citet{Kam07} 
differs from ambipolar diffusion-driven field decay. When ambipolar diffusion 
in the core - and the associated heating - is taken into account, the NS core 
can remain at fairly high temperatures for a long time \textit{if the 
decaying field is around $\sim 10^{16}$ G}.
\section{Conclusions}
\label{conclusions}
In this paper we have investigated some implications of one of the
key ansatz of the magnetar model; namely, that magnetars do form with 
millisecond spin periods and a (mainly) toroidal magnetic field, generated 
through the strong differential rotation of the collapsing proto-neutron 
star (\citealt{DT92} and \citealt{TD93}).

Building on our earlier work (\citealt{Ste05}, \citealt{DaSt}), we 
showed that one major implication of this scenario is that such 
objects can become strong sources of GWs in the first few days after 
formation. This results 
\textit{if} there exists at least a tiny misalignment (angle $\chi$) between 
the rotation axis and the symmetry axis of the magnetic field, at birth. 
The newly born NS is distorted to a prolate shape by the toroidal field, and 
is freely precessing because of $\chi \neq 0$. Under these circumstances 
internal viscous dissipation of the precessional motion will drive the 
magnetic symmetry axis orthogonal to the spin axis, the most favourable 
geometry for GW emission.

We discussed current uncertainties in various aspects of the model and
introduced simple approximations to treat each of them. 
We developed a simple analytical model describing the early rotational 
evolution of newly formed magnetars, that includes an ideal magnetic dipole 
torque plus the GW torque acting on an orthogonal, prolate rotator (see eq. 
\ref{pdot}). We then estimated the magnitude of GW emission 
from magnetars s as a function of their initial spin, internal magnetic 
energy and external (dipole) magnetic field (as well as their mass and 
radius). \\
Our main conclusions can summarized as follows:
\begin{itemize}

\item if magnetars are born with spin period less than 3 ms, 
internal toroidal fields $\geq 3 \times 10^{16}$ G and external dipole fields
$\leq 2 \times 10^{14}$ G then the expected GW signal 
would be strong enough to be detectable with Advanced 
LIGO/Virgo class detectors out to the Virgo cluster, where their formation
rate may be $\sim$ 1 per year; 

\item the estimated optimal S/N ratios for match-filtered signal 
searches (with one detector only) are very encouraging. However, as already 
noted
by \citet{Ste05}, optimal signal searches have unaffordable computational
costs. The developement of sub-optimal signal search strategies is  
required, a task that is currently under way.

\item if our scenario holds, the rotational energy $\simeq 3\times 10^{52}$ 
(P/ms)$^2$ ergs of the newly formed magnetar will be emitted mostly
as GWs, in the first few days after formation. As a consequence, Supernova 
Remnants (SNRs) surrounding evolved magnetars would not be expected to bear 
the signature of an excess energy injection ($> 10^{51}$ erg) soon after 
formation, since GWs do not interact appreciably with the 
expanding shell.

\end{itemize}

In particular, the condition that a newly formed magnetar be detectable 
as a GW source from Virgo-cluster distances turns out to be almost equivalent 
to the condition that it radiate less than $10^{51}$ erg through magnetic 
dipole radiation. The two requirements are met for nearly overlapping regions
of the internal and external magnetic fields parameter space. 
Stated differently, if the Galactic magnetar candidates studied by \citet{ViKu}
were born with millisecond spin periods, B$_t > 3 \times 10^{16}$ G and 
B$_d \leq 2 \times 10^{14}$, they would have lost most of their rotational 
energy through GW emission. Their SNRs would thus not show any excess energy 
compared to other SNRs - as observed, indeed - \textit{and} the GW signal 
they emitted could have been detected out to $\sim$ 20 Mpc with 
Advanced LIGO/Virgo class interferometers.
%

Finally, we considered the evolution of an internal field $> 10^{16}$ G
as a result of ambipolar diffusion, as already envisaged by \citet{GR92} 
and \citet{TD96}. Our aim here was twofold: first, we investigate this high
B-field regime, for which the calculations by \citet{TD96} are not appropriate. 
Second, we showed that even fields this strong have (at least) a slow decay 
mode through ambipolar diffusion, that is active 
soon after formation. This process can prevent the cooling of the magnetar 
core below a temperature of $\sim 10^9$ K for hundreds to thousands years. 
This conclusion is expected to have significant implications for our 
understanding of AXPs/SGRs.
%

\appendix

\section{Spin energy and freebody precession energy of a rotating ellipsoid}
\label{appendixa}
\begin{figure}
\centering
\includegraphics[width=3.0in, height=3.0in]{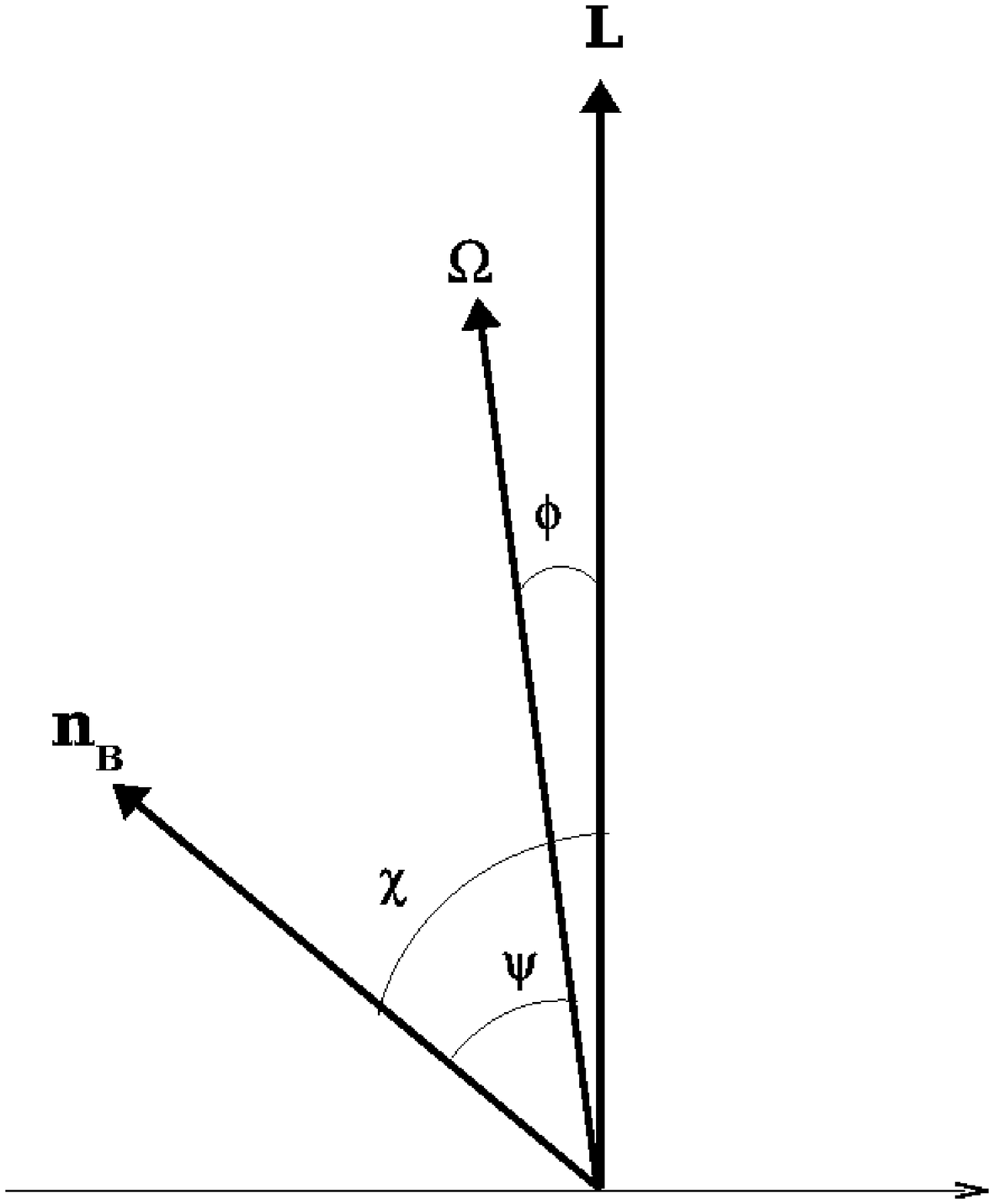}
\caption{Schematic representation of the angles of interest to the problem. 
L represents the invariant angular momentum vector, $\Omega$ is the angular 
frequency vector and n$_B$ is represents the unit vector along the simmetry 
axis of the magnetic field. The angle $\phi \simeq \epsilon_B \mbox{sin}\psi 
\mbox{cos}\psi \ll \psi$ \citep{JoAn01}. Therefore, $\psi 
\approx \chi$. }
\label{angles}
\end{figure}

We give here a quick derivation of the expression for the energy of freebody 
precession of a fluid star subject to both centrifugal and magnetic 
deformations. A more general discussion, in the context of ``ordinary'' NSs, 
is found in \citet{JoAn01}.\\
First of all, we write down the frequency of the freebody precession mode 
derived by \citet{MeTa72}
\begin{equation}
\label{precessionfrequency}
\omega_{\mbox{\tiny{pre}}} = \frac{I_3 - I_1}{I_1} \Omega \mbox{\small{cos}}
\psi = \epsilon_B \Omega \mbox{\small{cos}} \psi~~,
\end{equation}
where angles throughout this section are those defined in Fig. \ref{angles}.\\
Following \citealt{JoAn01} (and references therein), we define the
moment of inertia tensor of the fluid NS as a linear combination of three
contributions, namely a (spherical) gravitational part, plus two axisimmetric 
perturbations provided, in our case, by the centrifugal and magnetic fields, 
respectively. Hence:
\begin{equation}
\label{inertiatensor}
\underline{\mbox{\boldmath$I$}} = I_0 \underline{\mbox{\boldmath$\delta$}}
+ \Delta I_{\Omega} (\hat{\mbox{\boldmath$n$}}_{\Omega} 
\hat{\mbox{\boldmath$n$}}_{\Omega} - \underline{\mbox{\boldmath$\delta$}}/3) +
\Delta I_B (\hat{\mbox{\boldmath$n$}}_B 
\hat{\mbox{\boldmath$n$}}_B - \underline{\mbox{\boldmath$\delta$}}/3)
\end{equation}
where $\underline{\mbox{\boldmath$\delta$}}$ is the identity,
$\hat{\mbox{\boldmath$n$}}_{\Omega,B}$ represent the unit vectors along the 
spin and magnetic field axis and $\Delta I_{\Omega,B}$ are the magnitudes of 
the corresponding perturbations of the inertia tensor. The eigenvalues of the
latter are \citep{JoAn01}:
\begin{eqnarray}
\label{eigenvalues}
\mbox{\small{I}}_1 & = & \mbox{\small{I}}_0 + \frac{2}{3} \Delta 
\mbox{\small{I}}_{\Omega} - \frac{1}{3} \Delta \mbox{\small{I}}_B
\nonumber \\
\mbox{\small{I}}_2 & = & \mbox{\small{I}}_0 + \frac{2}{3} \Delta 
\mbox{\small{I}}_{\Omega} - \frac{1}{3} \Delta \mbox{\small{I}}_B 
\nonumber \\
\mbox{\small{I}}_3 & = & \mbox{\small{I}}_0 + \frac{2}{3} \Delta 
\mbox{\small{I}}_{\Omega} + \frac{2}{3} \Delta \mbox{\small{I}}_B =
\mbox{\small{I}}_1 + \Delta \mbox{\small{I}}_B 
\end{eqnarray}
These expressions show that the centrifugal deformation modifies all 
eigenvalues in the same way, as if it was effectively a spherically symmetric, 
additive term. 
The magnetic deformation, on the other hand, does introduce an asimmetry 
between the first two eigenvalues and the third one, inducing the magnetic 
ellipticity $\epsilon_B \equiv (\mbox{\small{I}}_3 - \mbox{\small{I}}_1) / 
\mbox{\small{I}}_1$.
The angle $\phi$ between $\mbox{\boldmath$\Omega$}$ and the angular momentum 
axis is always smaller than the tilt angle $\chi$ of the magnetic symmetry axis
\citep{JoAn01}. In formulae, to first order in $\epsilon_B$ it can
be shown that (by using eq. \ref{inertiatensor})
\begin{equation}
\label{smalltheta}
\mbox{\small{sin}} \phi \approx \phi \simeq \epsilon_B~
\mbox{\small{sin}} \psi \mbox{\small{cos}}\psi \ll \psi~~,
\end{equation}
from which $\psi = \chi - \phi \simeq \chi$.\\
Following the argument by \citet{CuJo01}, the NS angular momentum 
can be written as $\mbox{\boldmath$L$} = \underline{\mbox{\boldmath$I$}}
\mbox{\boldmath$\Omega$}$ and its kinetic energy as E$_{\mbox{\tiny{k}}} =
(1/2) \underline{\mbox{\boldmath$I$}} \mbox{\boldmath$\Omega$} \cdot 
\mbox{\boldmath$\Omega$}$, from which (to first order in $\epsilon_B$)
\begin{eqnarray}
\label{LandEk}
\mbox{\small{L}} & = &\mbox{\small{I}}_1 \Omega \left(1 + 2 \epsilon_B 
\mbox{\small {cos}}^2 \psi \right)^{1/2} \simeq \mbox{\small{I}}_1 \Omega 
\left(1 + 2 \epsilon_B \mbox{\small {cos}}^2 \chi \right)^{1/2} \nonumber \\
\mbox{\small{E}}_{\mbox{\tiny{k}}} & = & \frac{1}{2}
\mbox{\small{I}}_1 \Omega^2 (1 + \epsilon_B \mbox{\small{cos}}^2 \psi) \simeq 
\frac{1}{2} \mbox{\small{I}}_1 \Omega^2 (1 + \epsilon_B \mbox{\small{cos}}^2 
\chi) . 
\end{eqnarray}
From the latter equation we can obtain 
the energy of freebody precession by taking the difference between the 
second equation in (\ref{LandEk}) and the spin energy of the same 
ellipsoid that, with the same total angular momentum $\mbox{\boldmath$L$}$, 
spins around the axis having the greatest moment of inertia. The latter 
configuration is indeed the one that minimizes the energy, at constant angular 
momentum. It is thus the one towards which a freely precessing spheroid will 
evolve, given an internal dissipative process \citep{MeTa72, Jon76, Cut02}. 
Therefore, if $\Omega$ is the angular frequency of the precessing spheroid and 
$\Omega_{\mbox{\tiny{F}}}$ the angular frequency once freebody precession is 
completely damped, conservation of angular momentum implies
\begin{equation}
\label{conserveL}
\left(\frac{\Omega_{\mbox{\tiny{F}}}}{\Omega} \right)^2
= \left( \frac{\small{I}_1}{\small{I}_{\mbox{\tiny{max}}}}\right)^2
\left( 1 + 2\epsilon_B \mbox{\small{cos}}^2 \chi \right) .
\end{equation}
Writing the minimum spin energy as $\mbox{\small{E}}_
{\mbox{\tiny{min}}} = \frac{1}{2} \mbox{\small{I}}_{\mbox{\tiny{max}}} 
\Omega^2_{\mbox{\tiny{F}}}$, the freebody precession energy is  
\begin{eqnarray}
\label{wobble-energy}
\mbox{\small{E}}_{\mbox{\tiny{prec}}} & = & \mbox{\small{E}}_{\mbox{\tiny{k}}}
- \frac{1}{2} \mbox{\small{I}}_{\mbox{\tiny{max}}} \Omega^2_{\mbox{\tiny{F}}}  
\nonumber \\
 & = & \frac{1}{2} \mbox{\small{I}}_1 \Omega^2 \left \{ 1 + \epsilon_B  
\mbox{\small{cos}}^2 \chi - \frac{\mbox{\small{I}}_1}
{\mbox{\small{I}}_{\mbox{\tiny{max}}}} \left( 1 + 2 \epsilon_B\right)\right \} .
\end{eqnarray}
This general expression can be specialized to the case of an oblate ($
\mbox{\small{I}}_{\mbox{\tiny{max}}} = \mbox{\small{I}}_3$) or a prolate 
($\mbox{\small{I}}_{\mbox{\tiny{max}}} = \mbox{\small{I}}_1 = 
\mbox{\small{I}}_2$) ellipsoid giving\footnote{Note that the 
precession energy is positive in both cases, as it \textit{must} be}, 
respectively
\begin{eqnarray}
\label{precenergy}
\mbox{\small{E}}_{\mbox{\tiny{prec}}} & \simeq & ~~\frac{1}{2} 
\mbox{\small{I}}_1 \Omega^2 \epsilon_B \mbox{\small{ sin}}^2 \chi~~~
\mbox{\small{ oblate ellipsoid ~($\epsilon_B >0$)}} \nonumber \\
\mbox{\small{E}}_{\mbox{\tiny{prec}}} & \simeq & - \frac{1}{2} 
\mbox{\small{I}}_1 \Omega^2 \epsilon_B \mbox{\small{ cos}}^2 \chi~~
\mbox{\small{ prolate ellipsoid ($\epsilon_B < 0$)}}.
\end{eqnarray}
to first order in $\epsilon_B$.\\
Self consistence of  the above is warranted by $\mbox{\small{E}}_
{\mbox{\tiny{prec}}} \rightarrow 0$ for $\chi \rightarrow 0$ in the oblate 
case, and for $\chi \rightarrow \pi /2$ in the prolate case.\\
From the above formulae we can eventually relate the time derivative of the 
freebody precession energy to the time derivative of the tilt angle $\chi$, 
in the case of a prolate ellipsoid. 
Since conservation of angular momentum is required, $\Omega$ changes
only as a consequence of changes in $\chi$ (cfr. eq. \ref{conserveL}), which 
makes the time derivative of $\mbox{\small{E}}_{\mbox{\tiny{prec}}}$ a 
function of $\dot{\chi}$ only.
Taking the time derivatives of the first of eq. (\ref{LandEk}) and the 
second of eq. (\ref{precenergy}) and requiring angular momentum conservation
we obtain (to first order in $\epsilon_B$)
\begin{equation}
\label{A9}
\frac{\mbox{\small{d}} \mbox{\small{E}}_{\mbox{\tiny{prec}}}}
{\mbox{\small{dt}}} \simeq \mbox{\small{I}}_1 \epsilon_B \Omega^2 
\dot{\chi} \mbox{\small{cos}} \chi  \mbox{\small{sin}} \chi 
= -2 \mbox{\small{E}}_{\mbox{\tiny{prec}}} \tau^{\mbox{\tiny{$\chi$}}}_{\mbox{\tiny{d}}} 
\end{equation}
where $\tau^{\mbox{\tiny{$\chi$}}}_{\mbox{\tiny{d}}}$ was defined in $\S$ 
\ref{bulk}.
\section{Calculation of the damping time of freebody precession through bulk
viscosity} 
\label{appendixb}
In this appendix we describe the calculation that leads to the estimated 
timescale for dissipation of the free precessional motion through bulk 
viscosity (eq. \ref{taudamp}).
According to the definition of $\tau_{\mbox{\tiny{d}}}$ (eq. \ref{deftau}), we
need an expression for both the bulk viscosity coefficient, $\zeta$, and
the precession-induced density perturbation, $\delta \rho$.

For the bulk viscosity coefficient, we recall here the general
expression (\ref{bulkgeneral})
\begin{equation}
\label{bulkgenappendix}
\mbox{Re}(\zeta) = \frac{n \tau \left(\partial p/ \partial x\right)_n
d\tilde{x}/dn }{1+ (\omega \tau)^2}
\simeq  \frac{n \left(\partial p/ \partial x\right)_n d\tilde{x}/dn}
{\omega^2 \tau} .
\end{equation}
%
For $npe$ matter, as we have assumed throughout, the timescale for
$\beta$-reactions ($\tau_{\beta}$) is (cfr. \citealt{RG92})
\begin{equation}
\label{tau}
\tau_{\beta} = \frac{3 n_p}{\lambda_{\beta} E_{F_n}} \simeq \frac{0.23}{T^6_9} \left(
\frac{\rho}{\rho_{\mbox{\tiny{nuc}}}}\right)^{\frac{2}{3}}~~\mbox{yr} . 
\end{equation}
Here $\tilde{x}$ is the equilibrium fraction of charged particles (see below), 
$E_{F_n} = (h^2/2m_n)(3 \pi^2 n_n)^{2/3}$ is the neutron Fermi energy and 
$\lambda_{\beta} \simeq 5 \times 10^{33} T^6_9 (\rho/\rho_{\mbox{\tiny{nuc}}})^{2/3} $ 
erg$^{-1}$ s$^{-1}$ is the rate of $\beta$-reactions when the combined fluid is 
out of chemical equilibrium by an amount $\delta \mu = \delta \mu_n - \delta 
\mu_e - \delta \mu_p$. Finally, $\rho_{\mbox{\tiny{nuc}}} \approx 2.8 \times 
10^{14}$ g cm$^{-3}$ is the nuclear saturation density.
The last step in eq. (\ref{bulkgeneral}) holds for $\omega_{\mbox{\tiny{pre}}} 
\tau_{\beta} \gg 1$, the relevant approximation here.\\
The pressure of degenerate $npe$ matter, neglecting the small contribution 
from protons, is given by $
p \simeq \frac{2}{5} m_n E_{F_n} + \frac{1}{4} m_e E_{F_e}
%
$
where the equilibrium proton (and electron) fraction is \citep{RG92}
\begin{equation}
\label{x}
\tilde{x} = \left(\frac{n_p}{n_n}\right)_{\mbox{\tiny{eq}}} \approx \frac{n_p}{n} 
\simeq 6 \times 10^{-3} \frac{\rho}{\rho_{\mbox{\tiny{nuc}}}} .
\end{equation}
From all the above, one obtains the required expression (cfr. \citealt{Saw89}): 
\begin{eqnarray}
\label{bulkcoefficient}
\zeta & \approx & 6 \times 10^{-59} \frac{\rho^2 T^6}
{\omega_{\mbox{\tiny{pre}}}^2}  \approx  \nonumber \\
& \approx  & \frac{5.6 \times 10^{29}}{\mbox{\small{cos}}^2 \chi}
\left(\frac{\rho_{15}}{0.7}\right)^2\left(\frac{\mbox{\small{P}}}
{\mbox{\small{ms}}}\right)^2 \left(\frac{10^{50}\mbox{\small{erg}}}
{\mbox{\small{E}}_{\mbox{\tiny{B}}}}\right)^2 T^6_{10} \frac{\mbox{\small{erg}}}
{\mbox{\small{cm}}} 
\end{eqnarray}
As a starting poin to determine the density perturbation $\delta \rho$, we 
refer to the expression given by \citet{MeTa72} of the centrifugal 
distortion of a fluid star, $\rho_{\Omega}$, as a function of a spherical 
coordinate system whose origin is at the star center and whose pole is the 
magnetic pole $(r, \theta, \lambda)$ .\\
The amplitude of internal field of motion, $\xi$, is determined by the 
non-spherical part of the rotational distortion of the fluid. 
The perturbation of the density profile is
\begin{equation}
\label{appendeltaro}
\delta \rho_{\Omega} = \rho_{\Omega} (\lambda -\Omega t) - \rho_{\Omega} 
(\lambda) = \frac{1}{2} f(r) \hat{K}(\chi, \theta, \lambda, \Omega t)~~.
\end{equation}
where the function $\hat{K} (\chi, \theta, \lambda, \Omega)$ is
\begin{eqnarray}
\label{K}
\hat{K}(\chi, \theta, \lambda, \Omega) & = & \mbox{sin}^2 \chi[1 - P_2(\mu)] 
\times \nonumber \\
 & \times & \left\{\mbox{sin} 2\lambda \mbox{ sin} 2\Omega t - \mbox{cos} 
2\lambda (1- \mbox{cos} 2 \Omega t)\right\} + \nonumber \\
 & + & 3 \mbox{ sin} \chi \mbox{cos} \chi \mbox{sin} 2\theta~[\mbox{sin} 
\lambda \mbox{ sin} \Omega t - \nonumber \\
& - & \mbox{cos}\lambda (1 -\mbox{cos}\Omega t)]~~.
\end{eqnarray}
Here, $P_2(\mu) = P_2(\mbox{\small{cos}}\theta)$ is the Legendre polynomial. 
The radial function $f(r)$, that depends on the stellar model 
assumed, can only be obtained analytically through approximate (and/or 
idealized) calculations. \\
In particular, the centrifugal deformation of a fluid mass with a polytropic 
equation of state is an old problem that has been largely discussed in the 
literature. We refer, here, to the study by \citet{Cha33}, where both 
analytical formulae and numerical values for the relevant functions are 
derived.\\
Given the polytropic EOS $p = k \rho^{1+1/n}$, one can define an 
adimensional density variable
\begin{equation}
\label{polyt1}
\rho = \rho_c \left(\frac{\rho}{\rho_c}\right)^n = \rho_c \Theta^n~.
\end{equation}
The ratio of the rotation energy to the gravitational binding energy defines
the rotation parameter $v$
\begin{equation}
\label{rotate}
v = \frac{\Omega^2}{2 \pi \mbox{\small{G}} \rho_c}~~.
\end{equation}
in powers of which the adimensional density $\Theta$ can be expanded. 
In the above G is the gravitational constant and $\rho_c$ the central density 
of the fluid star. For $v \ll 1$, a first order expansion will suffice.
Introducing the adimensional radial coordinate $\xi$, such that $r = \alpha 
\xi$, the first order expansion of the density profile becomes
\begin{equation}
\label{chandrazero}
\Theta(\xi) = \theta_0(\xi)  + v \psi(\xi)~~,
\end{equation}
where $\theta_0$ is the unperturbed density profile, solution to the 
Lane-Emden equation for a non-rotating polytrope.
%
%
The parameter $\alpha$ above is related to the polytropic equation of state
\begin{equation}
\label{alfapolyt}
\alpha = \left[\frac{(n+1)~k}{4 \pi \mbox\small{G}}~\rho_c^{-1+1/n}\right]^
{1/2} \stackrel{(n=1)}{\rightarrow} \left(\frac{k}{2 \pi \mbox{\small{G}}} 
\right)^{1/2}~~.
\end{equation}
The problem is thus reduced to finding an appropriate expression for the 
function $\psi(\xi)$. In particular, for polytropes of index $n$ 
the following expansion holds
\begin{equation}
\label{chandrauno}
\Theta (\xi) = \theta_0 + v\left[ \psi_0 (\xi) + A_2 \psi_2(\xi) P_2(\mu)
\right]~~,
\end{equation}
where $\psi_0(\xi)$ is the spherical part of the centrifugal deformation and
the non-spherical part of the deformation corresponds to the second term in
square parenthesis. The coefficient $A_2$ is
\begin{equation}
\label{chandradue}
A_2 = - \frac{5}{6} \frac{\xi^2_R}{3 \psi_2(\xi_R) + \xi_R \psi^{'}_2(\xi_R)}~~,
\end{equation}
where $\alpha \xi_R = R$ and the prime denotes a first derivative. An 
approximate analytical expression for $\psi_2(\xi)$ is given as an 
eighth-order polynomial expansion in $\xi$ \citep{Cha33}
\begin{equation}
\label{psidue}
\psi_2(\xi) = \xi^2 - \frac{1}{14} \xi^4 + \frac{1}{504} \xi^6 - 
\frac{1}{33264}\xi^8 + \mbox{\textit{o}}(\xi^{10})~~,
\end{equation}
which formally completes the problem at hand.\\
As it has been shown by numerical integration of the equilibrium equations,
a first-order expansion in $v$ provides a relatively good approximation 
to slowly rotating polytropes; in the case $n=1$, $v \leq 0.075$ gives 
approximately the rotation rate at which deviations from the first order 
perturbation theory become non-negligible \citep{Tas78}. In particular,
direct integration of the equilibrium equations reveals a systematically 
larger deformation of rapid rotators compared to the results obtained through 
the first order perturbative approximation. Therefore, our general expectation 
would be that the latter underestimates the real deformation of a rapidly
rotating NS ($\delta \rho$), thus underestimating the efficiency of bulk 
viscous dissipation.\\
The polytropic index $n=1$ greatly simplifies the analytical treatment and
allows one to derive quite straightforward formulae. The adimensional radius 
of the polytropic star will be $\xi _R = \pi$ and the numerical values of 
$\psi_2(\xi_R)$ and $\psi^{'}_2(\xi_R)$ have been tabulated by \citet{Cha33}, 
allowing to derive $A_2 (n=1) \simeq -0.54833$. Finally, $\alpha$ is 
determined by the linear scale for the NS radius so that:
\begin{equation}
\label{alfavalue}
\alpha \simeq 3.82 \times 10^5 \left(\frac{\mbox{\small{R}}}
{12~\mbox{\small{Km}}}\right)~~\mbox{\small{cm}}
\end{equation}
and the EOS thus becomes $p =6.1146 \times 10^4 \left(\mbox{\small{R}}
/ 12~\mbox{\small{Km}}\right)^2 \rho^2$.\\
The density perturbation of eq. (\ref{appendeltaro}), the one that determines 
the precession-induced internal motions damped by bulk viscosity, amounts to 
only the non-spherical part of the centrifugal deformation. The spherical
part does not cause any radial periodic motions and, thus, does not perturb 
the local chemical potential equilibrium of NS matter. We are therefore 
left with:
\begin{equation}
\label{correspondence}
f(r(\xi)) = \rho_c v A_2 \psi_2(\xi)
\end{equation}
where the angular part corresponds to the function $K(\chi, \theta, \lambda,
\Omega)$. This must be inserted in eq. (\ref{appendeltaro}) to obtain the 
full expression for the non-spherical density perturbation. Recalling 
eq. (\ref{deftau}), we thus get the full expression for the bulk viscosity 
damping rate:
\begin{equation}
\label{dampingbulk}
\dot{\mbox{E}}_{\mbox{\tiny{diss}}} = 60T^{6}_{10}~\omega^2 \int 
\mbox{\small{dV}} \left(\frac{\delta \rho}{\rho}\right)^2 \frac{\rho^2}
{\omega^2} = 60T^{6}_{10}\int \mbox{\small{dV}} (\delta \rho)^2
\end{equation}
where:
\begin{eqnarray}
\label{express}
[\delta \rho(r(\xi))]^2 & = & \frac{1}{4} f^2(r(\xi)) 
K^2(\chi, \theta, \lambda, \Omega) =\nonumber \\
 & = & \frac{1}{4} (\rho_c v)^2 A^2_2 \psi^2_2(\xi) \hat{K}^2(\chi, \theta, 
\lambda, \Omega) 
\end{eqnarray}
so that, eventually, the full expression for $\dot{\mbox{E}}_{\mbox{\tiny{diss}}}$ 
can be written as:
\begin{eqnarray}
\label{dissip}
\dot{\mbox{E}}_{\mbox{\tiny{diss}}} & \approx & 2.6 \times 10^{13}
T^{6}_{10} \Omega^4 \alpha^3 \int \psi^2_2(\xi) \xi^2 \mbox{\small{d}}\xi 
\int \hat{K}^2
\mbox{\small{ sin}}\theta 
\mbox{\small{d}}\theta\mbox{\small{d}}\lambda  \nonumber \\
 & \approx & 2.2 \times 10^{45} \left(\frac{\mbox{\small{T}}}{10^{10}
\mbox{\small{K}}}\right)^{6} \left(\frac{\mbox{\small{ms}}}{\mbox{\small{P}}}
\right)^4\left(\frac{\mbox{\small{R}}}{12\mbox{\small{Km}}}\right)^3 \times
\nonumber \\
 & \times & \int \psi^2_2(\xi) \xi^2 \mbox{\small{d}}\xi \int \hat{K}^2 
\mbox{\small{sin}}\theta \mbox{\small{d}}\theta \mbox{\small{d}}\lambda 
\end{eqnarray}
The radial integral, solved from $\xi=0$ to $\xi_R=\pi$ gives 138.04, while 
the angular part gives $(24 \pi/5) \mbox{\small{sin}}^2 \chi 
(1+3 \mbox{\small{cos}}^2\chi)$. Eventually:
\begin{eqnarray}
\label{dissipfinal}
\dot{\mbox{E}}_{\mbox{\tiny{diss}}}  & \approx  & 4.7 \times 10^{48}
\mbox{\small{ sin}}^2 \chi (1+3 \mbox{\small{cos}}^2\chi) \nonumber \\
& & \left(\frac{\mbox{\small{T}}}{10^{10}
\mbox{\small{K}}}\right)^{6} \left(\frac{\mbox{\small{ms}}}
{\mbox{\small{P}}}\right)^4
\left(\frac{\mbox{\small{R}}}{12\mbox{\small{Km}}}\right)^3 
\mbox{\small{erg s}}^{-1} 
\end{eqnarray}
As a last step, given the energy of the freebody precession mode derived in 
$ \S$ \ref{appendixa}, we obtain the dissipation timescale (see eq. 
\ref{taudamp}):
\begin{eqnarray}
\label{definitive}
\tau_{\mbox{\tiny{d}}} =  \frac{2E_{\mbox{\tiny{pre}}}}
{\dot{\mbox{E}}_{\mbox{\tiny{diss}}}} & \simeq & \frac{13.5~\mbox{\small{cos}}^2 
\chi}{\mbox{\small{sin}}^2 \chi (1+3 \mbox{\small{cos}}^2 \chi)} 
\left(\frac{\mbox{\small{E}}_B}{10^{50}{\mbox{\small{erg}}}}\right)^2
\left(\frac{\mbox{\small{P}}}{\mbox{\small{ms}}}\right)^2
\nonumber \\
& \times & \left(\frac{\mbox{\small{T}}}{10^{10}{\mbox{\small{K}}}}\right)^{-6} 
\left(\frac{\mbox{\small{M }}}{\mbox{\small{1.4M}}_{\odot}}\right) 
\left(\frac{\mbox{\small{R}}}{12{\mbox{\small{Km}}}}\right)^3
\mbox{\small{s}}
\end{eqnarray}

\section*{Acknowledgments}
This work was supported by Virgo-EGO Scientific Forum (VESF) Fellowship
at the University of Pisa. \\
S.D. acknowledges N. Andersson for suggesting to check for the possible 
role of bulk viscosity in damping freebody precession of a newly formed NS.\\
As one of the thousands Italian researchers with medium-term positions, SD 
acknowledges the support of Nature (455, 835-836) and thanks the Editors for 
increasing the international awareness of the current critical situation of 
Italian Research and of young researchers \textit{in primis}.


\begin{thebibliography}{99}
\bibitem[\protect\citeauthoryear{Abbot et al.}{2007}]{Abb07} 
Abbott, B., et al., 2007, PRD, 76, 042001 
\bibitem[\protect\citeauthoryear{Allen \& Horvath}{2004}]{AlHo}  Allen, M. P., 
\& Horvath, J. E.\ 2004, Ap.J., 616, 346 
\bibitem[\protect\citeauthoryear{Alpar \& Sauls}{1988}]{AlSa88}
Alpar, M. A., \& Sauls, J. A., 1988, Ap.J., 327, 723
\bibitem[\protect\citeauthoryear{Arons}{2003}]{Ar03} Arons, J., 2003, Ap.J. 
589, 871 
\bibitem[\protect\citeauthoryear{Arras et al.}{2004}]{Arr04} Arras, P., 
Cumming, A., \& Thompson, C., 2004, Ap.J.L., 608, L49 
\bibitem[\protect\citeauthoryear{Blandford, Applegate \& Hernquist}{1983}]
{BlApHe83} Blandford, R. D., Applegate, J. H., \& Hernquist, L., 1983, 
M.N.R.A.S., 204, 1025 
\bibitem[\protect\citeauthoryear{Blandford \& Romani}{1988}]{BlaRo88}
 Blandford, R. D., \& Romani, R. W., 1988, M.N.R.A.S., 234, 57P 
\bibitem[\protect\citeauthoryear{Bonazzola \& Gourgoulhon}{1996}]{BoGou} 
Bonazzola, S., \& Gourgoulhon, E. 1996, Astr. \& Astroph., 312, 675
\bibitem[\protect\citeauthoryear{Bucciantini et al.}{2006}]{Buc06}
Bucciantini, N., Thompson, T. A., Arons, J., Quataert, E., 
\& Del Zanna, L., 2006, M.N.R.A.S., 368, 1717
\bibitem[\protect\citeauthoryear{Bucciantini et al.}{2007}]{Buc07}
Bucciantini, N., Quataert, E., Arons, J., Metzger, B. D., \& Thompson, 
T. A., 2007, M.N.R.A.S., 380, 1541 
\bibitem[\protect\citeauthoryear{Camilo et al.}{2000}]{Cam00}
Camilo, F., Kaspi, V. M., Lyne, A. G., Manchester, R. N., Bell, J. F., 
D'Amico, N., McKay,N. P. F., \& Crawford, F., 2000, Ap.J., 541, 367 
\bibitem[\protect\citeauthoryear{Chandrasekhar}{1933}]{Cha33} Chandrasekhar, S. 
1933, M.N.R.A.S., 93, 390 
\bibitem[\protect\citeauthoryear{Contopoulos, Kazanas \& Fendt}{1999}]{CoKaFe99}
Contopoulos, I., Kazanas, D., \& Fendt, C., 1999, Ap.J., 511, 351 
\bibitem[\protect\citeauthoryear{Contopoulos \& Spitkovsky}{2006}]{CoSp06}
Contopoulos, I., \& Spitkovsky, A., 2006, Ap.J., 643, 1139 
\bibitem[\protect\citeauthoryear{Cumming et al.}{2004}]{Cu04} Cumming, A., 
Arras, P., \& Zweibel, E., 2004, Ap.J., 609, 999 
\bibitem[\protect\citeauthoryear{Cutler \& Jones}{2001}]{CuJo01}
Cutler, C., \& Jones, D. I., 2001, PRD, 63, 024002 
\bibitem[\protect\citeauthoryear{Cutler}{2002}]{Cut02} Cutler, C., 2002, PRD, 
66, 084025 
\bibitem[\protect\citeauthoryear{Dall'Osso et al.}{2003}]{Dal03}
Dall'Osso, S., Israel, G. L., Stella, L., Possenti, A., \& Perozzi, E., 2003, 
Ap.J., 599, 485 
\bibitem[\protect\citeauthoryear{Dall'Osso \& Stella}{2007}]{DaSt} Dall'Osso, 
S., \& Stella, L., 2007, Astroph. \& Space Science, 308, 119 
\bibitem[\protect\citeauthoryear{Dib et al.}{2008}]{Dib08}
Dib, R., Kaspi, V. M., \& Gavriil, F. P., 2008, Ap.J., 673, 1044 
\bibitem[\protect\citeauthoryear{Duncan}{1998}]{Du98} Duncan, R. C. 1998, 
Ap.J.L., 498, L45 
\bibitem[\protect\citeauthoryear{DT92}{}]{DT92} Duncan, R. C., \& Thompson, C. 
1992, Ap.J.L., 392, L9 
\bibitem[\protect\citeauthoryear{Ferrario \& Wickramasinghe}{2006}]{FerWick}
Ferrario, L., \& Wickramasinghe, D. 2006, M.N.R.A.S., 367, 1323 
\bibitem[\protect\citeauthoryear{Gaensler et al.}{1999}]{Gae99} Gaensler, 
B. M., Gotthelf, E. V., \& Vasisht, G., 1999, Ap.J.L., 526, L37 
\bibitem[\protect\citeauthoryear{Geppert \& Rheinhardt}{2006}]{GepRe}
Geppert, U., \& Rheinhardt, M., 2006, Astr. \& Astroph., 456, 639 
\bibitem[\protect\citeauthoryear{Goldreich}{1970}]{Gol70} Goldreich, P., 
1970, Ap.J.L., 160, L11 
\bibitem[\protect\citeauthoryear{GR92}{}]{GR92} Goldreich, P., \& 
Reisenegger, A. 1992, Ap.J., 395, 250 
\bibitem[\protect\citeauthoryear{Gruzinov}{2006}]{Gru06}
Gruzinov, A., 2006, arXiv:astro-ph/0604364 
\bibitem[\protect\citeauthoryear{Haskell et al.}{2008}]{Has08}
Haskell, B., Samuelsson, L., Glampedakis, K., \& Andersson, N., 2008, 
M.N.R.A.S., 385, 531 
\bibitem[\protect\citeauthoryear{Ioka}{2001}]{Io01} Ioka, K., 2001, 
M.N.R.A.S., 327, 639
\bibitem[\protect\citeauthoryear{Israel et al.}{2007}]{Isr07}
Israel, G. L., G{\"o}tz, D., Zane, S., Dall'Osso, S., Rea, N., \& Stella, L., 
2007, Astr. \& Astroph., 476, L9  
\bibitem[\protect\citeauthoryear{Jones}{1976}]{Jon76}
Jones, P.~B.\ 1976, Astrophysics and Space Science, 45, 369 
\bibitem[\protect\citeauthoryear{Jones \& Andersson}{2001}]{JoAn01}
Jones, D. I., \& Andersson, N., 2001, M.N.R.A.S., 324, 811
\bibitem[\protect\citeauthoryear{Kaminker et al.}{2007}]{Kam07}
Kaminker, A. D., Yakovlev, D. G., Potekhin, A. Y., Shibazaki, N., Shternin, 
P. S., \& Gnedin, O. Y., 2007, Astroph. \& Space Science, 308, 423 
\bibitem[\protect\citeauthoryear{Lattimer \& Prakash}{2001}]{LaPr01} Lattimer, 
J. M., \& Prakash, M. 2001, Ap.J., 550, 426
\bibitem[\protect\citeauthoryear{Lattimer \& Prakash}{2007}]{LaPr07}
Lattimer, J. M., \& Prakash, M., 2007, Phys. Rep., 442, 109 
\bibitem[\protect\citeauthoryear{Lazzati et al.}{2005}]{Laz} Lazzati, D., 
Ghirlanda, G., \& Ghisellini, G., 2005, M.N.R.A.S., 362, L8 
\bibitem[\protect\citeauthoryear{Lindblom, Owen \& Morsink}{1998}]{LiOwMo98} 
Lindblom, L., Owen, B. J., \& Morsink, S. M., 1998, Phys. Rev. Lett., 80, 4843 
\bibitem[\protect\citeauthoryear{Lindblom, Mendell \& Owen}{1999}]{LiMeOw99} 
Lindblom, L., Mendell, G., \& Owen, B. J., 1999, PRD, 60, 064006 
\bibitem[\protect\citeauthoryear{Lindblom \& Owen}{2002}]{LiOw02} 
Lindblom, L., \& Owen, B. J. 2002, PRD, 65, 063006 
\bibitem[\protect\citeauthoryear{Link, Epstein \& Baym}{1992}]{LiEpBa92}
Link, B., Epstein, R. I., \& Baym, G., 1992, Ap.J.L., 390, L21
\bibitem[\protect\citeauthoryear{Livingstone et al.}{2007}]{Liv07}
Livingstone, M. A., Kaspi, V. M., Gavriil, F. P., Manchester, R. N., Gotthelf, 
E. V. G., \& Kuiper, L., 2007, Astroph. \& Space Science, 308, 317
\bibitem[\protect\citeauthoryear{Losurdo}{2007}]{Los07}
Losurdo G., 2007, ``Advanced Virgo sensitivity curve: cavity finesse and signal
recycling tuning'', Virgo Internal Report: VIR–024A–07
\bibitem[\protect\citeauthoryear{Lyne, Pritchard \& Smith}{1988}]{LyPrSm88}
Lyne, A. G., Pritchard, R. S., \& Smith, F. G., 1988, M.N.R.A.S., 233, 667
\bibitem[\protect\citeauthoryear{Lyne et al.}{1996}]{Lyn96}
Lyne, A. G., Pritchard, R. S., Graham-Smith, F., \& Camilo, F., 1996, Nature, 
381, 497 
\bibitem[\protect\citeauthoryear{Lyutikov}{2003}]{Lyu03} Lyutikov, M., 2003, 
M.N.R.A.S., 346, 540 
\bibitem[\protect\citeauthoryear{Marshall et al.}{2004}]{Mar04}
Marshall, F. E., Gotthelf, E. V., Middleditch, J., Wang, Q. D., 
\& Zhang, W., 2004, Ap.J., 603, 682 
\bibitem[\protect\citeauthoryear{Mazets et al.}{1979}]{Maz} Mazets, E. P., 
Golentskii, S. V., Ilinskii, V. N., Aptekar, R. L., \& Guryan, I. A. 1979, 
Nature, 282, 587 
\bibitem[\protect\citeauthoryear{Mereghetti \& Stella}{1995}]{MeSte} 
Mereghetti, S., \& Stella, L., 1995, Ap.J.L., 442, L17 
\bibitem[\protect\citeauthoryear{Mestel \& Takhar}{1972}]{MeTa72} Mestel, L., 
\& Takhar, H. S. 1972, M.N.R.A.S., 156, 419
\bibitem[\protect\citeauthoryear{Metzger et al.}{2007}]{Metz07} Metzger, 
B. D., Thompson, T. A., \& Quataert, E., 2007, Ap.J., 659, 561 
\bibitem[\protect\citeauthoryear{Middleditch et al.}{2006}]{Mid06}
Middleditch, J., Marshall, F. E., Wang, Q. D., Gotthelf, E. V., \& Zhang, W., 
2006, Ap.J., 652, 1531 
\bibitem[\protect\citeauthoryear{Nakar et al.}{2006}]{Nak}Nakar, E., Gal-Yam, 
A., Piran, T., \& Fox, D. B., 2006, Ap.J., 640, 849 
\bibitem[\protect\citeauthoryear{Ostriker \& Gunn}{1969}]{OstrGun} Ostriker, 
J. P., \& Gunn, J. E. 1969, Ap.J., 157, 1395
\bibitem[\protect\citeauthoryear{Owen et al.}{1998}]{Ow98}
Owen, B. J., Lindblom, L., Cutler, C., Schutz, B. F., Vecchio, A., 
\& Andersson, N., 1998, PRD, 58, 084020  
\bibitem[\protect\citeauthoryear{Owen \& Lindblom}{2002}]{OwLi02} 
Owen, B. J., \& Lindblom, L., 2002, Classical and Quantum Gravity, 19, 1247
\bibitem[\protect\citeauthoryear{Paczynsky}{1992}]{Pac} Paczynski, B., 1992,
Acta Astronomica, 42, 145 
\bibitem[\protect\citeauthoryear{Page et al.}{2004}]{Pag04} Page, D., 
Lattimer, J. M., Prakash, M., \& Steiner, A. W. 2004, Ap.J. Suppl., 155, 623
\bibitem[\protect\citeauthoryear{Page, Geppert \& Weber}{2006}]{Pag06}
Page, D., Geppert, U., \& Weber, F., 2006, Nuclear Physics A, 777, 497 
\bibitem[\protect\citeauthoryear{Palomba}{2001}]{Pal} Palomba, C., 2001, 
Astr. \& Astroph., 367, 525 
\bibitem[\protect\citeauthoryear{Pons \& Geppert}{2007}]{PoGe07}
Pons, J. A., \& Geppert, U., 2007, Astr. \& Astroph, 470, 303 
\bibitem[\protect\citeauthoryear{Pons et al.}{2007}]{Pon07}
Pons, J. A., Link, B., Miralles, J. A., \& Geppert, U., 2007, Phys. Rev. 
Lett., 98, 071101 
\bibitem[\protect\citeauthoryear{Popov \& Stern}{2006}]{PoSt}
Popov, S. B., \& Stern, B. E., 2006, M.N.R.A.S., 365, 885 
\bibitem[\protect\citeauthoryear{Reisenegger \& Goldreich}{1992}]{RG92} 
Reisenegger, A., \& Goldreich, P. 1992, Ap.J., 395, 240 
\bibitem[\protect\citeauthoryear{Ruderman}{1991}]{Rud91} Ruderman, M., 1991, 
Ap.J., 366, 261 
\bibitem[\protect\citeauthoryear{Ruderman, Zhu \& Chen}{1998}]{RuZhCh98}
Ruderman, M., Zhu, T., \& Chen, K., 1998, Ap.J., 492, 267 
\bibitem[\protect\citeauthoryear{Sawyer}{989}]{Saw89} Sawyer, R. F., 1989, 
PRD, 39, 3804 
\bibitem[\protect\citeauthoryear{Spitkovsky}{2006}]{Spi06}
Spitkovsky, A., 2006, Ap.J.L., 648, L51 
\bibitem[\protect\citeauthoryear{Stella et al.}{2005}]{Ste05} Stella, L., 
Dall'Osso, S., Israel, G. L., \& Vecchio, A., 2005, Ap.J.L., 634, L165 
\bibitem[\protect\citeauthoryear{Tanvir et al.}{2005}]{Tanv}
Tanvir, N. R., Chapman, R., Levan, A. J., \& Priddey, R. S., 2005, Nature, 
438, 991
\bibitem[\protect\citeauthoryear{Tassoul}{1978}]{Tas78}
Tassoul, J. L., 1978, Theory of Rotating Stars, Princeton University Press,
Princeton University, New Jersey
\bibitem[\protect\citeauthoryear{Thompson et al.}{2004}]{Tho04}
Thompson, T. A., Chang, P., \& Quataert, E., 2004, Ap.J., 611, 380 
\bibitem[\protect\citeauthoryear{TD93}{}]{TD93} Thompson, C., \& Duncan, R. C. 
1993, Ap.J., 408, 194 
\bibitem[\protect\citeauthoryear{TD95}{}]{TD95} Thompson, C., \& Duncan, R. C. 
1995, M.N.R.A.S., 275, 255 
\bibitem[\protect\citeauthoryear{TD96}{}]{TD96}
Thompson, C., \& Duncan, R. C. 1996, Ap.J., 473, 322 
\bibitem[\protect\citeauthoryear{TD01}{}]{TD01} Thompson, C., \& Duncan, R. C. 
2001, Ap.J., 561, 980 5 
\bibitem[\protect\citeauthoryear{Ushomirsky, Cutler \& Bildsten}{2000}]
{UsCuBi00} Ushomirsky, G., Cutler, C., \& Bildsten, L. 2000, M.N.R.A.S., 
319, 902 
\bibitem[\protect\citeauthoryear{Usov}{1992}]{Uso} Usov, V. V., 1992, Nature, 
357, 472  
\bibitem[\protect\citeauthoryear{Vink \& Kuiper}{2006}]{ViKu} Vink, J., \& 
Kuiper, L. 2006, M.N.R.A.S., 370, L14 
\bibitem[\protect\citeauthoryear{Woods \& Thompson}{2006}]{WoTh06}
Woods, P. M., \& Thompson, C., 2006, Compact stellar X-ray sources, 547 
\end{thebibliography}
\end{document}